\documentclass[pdflatex,sn-mathphys-num,default,iicol]{sn-jnl}

\usepackage{graphicx}%
\usepackage{multirow}%
\usepackage{amsmath,amssymb,amsfonts}%
\usepackage{amsthm}%
\usepackage{mathrsfs}%
\usepackage[title]{appendix}%
\usepackage{xcolor}%
\usepackage{textcomp}%
\usepackage{manyfoot}%
\usepackage{booktabs}%
\usepackage{algorithm}%
\usepackage{algorithmicx}%
\usepackage{algpseudocode}%
\usepackage{listings}%
\usepackage{nicematrix}
\usepackage[normalem]{ulem}
\usepackage{threeparttable} 
\usepackage{bm}

\theoremstyle{thmstyleone}%
\theoremstyle{thmstyletwo}%

\theoremstyle{thmstylethree}%

\begin{document}

\title[Effects of Stochastic Games on Evolutionary Dynamics in Structured Populations]{Effects of Stochastic Games on Evolutionary Dynamics in Structured Populations}

\author[1]{\fnm{Yuji} \sur{Zhang}}

\author*[1]{\fnm{Minyu} \sur{Feng}}\email{myfeng@swu.edu.cn}

\author*[2]{\fnm{Qin} \sur{Li}}\email{qinli1022@swu.edu.cn}

\author[3,4,5,6]{\fnm{Matja\v{z}} \sur{Perc}}

\author[7]{\fnm{Attila} \sur{Szolnoki}}

\affil[1]{College of Artiﬁcial Intelligence, Southwest University, Chongqing 400715, China}
\affil[2]{Business College, Southwest University, Chongqing 402460, China}
\affil[3]{Faculty of Natural Sciences and Mathematics, University of Maribor, 2000 Maribor, Slovenia}
\affil[4]{Community Healthcare Center Dr.~Adolf Drolc Maribor, 2000 Maribor, Slovenia}
\affil[5]{Department of Physics, Kyung Hee University, Seoul 02447, Republic of Korea}
\affil[6]{University College, Korea University, Seoul 02841, Republic of Korea}
\affil[7]{Institute of Technical Physics and Materials Science, Centre for Energy Research, 1525 Budapest, Hungary}

\abstract{Continuously changing environments have a paramount role in the evolution of cooperative behavior. Previous works have shown that the transitions among different games, as the feedback between behaviors and environments, can promote cooperative behavior significantly. Quantitative analysis, however, is limited to homogeneous populations, while realistic populations in nature are often more complex and highly heterogeneous. We hereby provide an analytical treatment of when the evolution of cooperation can be supported in stochastic games, applying to arbitrary spatial heterogeneity and payoff structure. We highlight that the rule and frequency of game changes can have surprisingly diverse effects on evolutionary outcomes, depending on the governing social dilemmas. While stochastic games favor the evolution of cooperation in donation games, this is not the case for public goods games and snowdrift games. Hence, our framework and model results offer a more subtle insight into the long-standing enigma.}

\maketitle

\section{Introduction}
Prosocial behavior, featured by collective cooperation, constitutes a fundamental component of social prosperity and development, which facilitates the mitigation of climate challenges~\cite{ferrari2023no}, the security of global food supplies~\cite{world2022global}, the scientific breakthrough in the study of cultural evolution~\cite{brewer2017grand}, and so forth. Nevertheless, a comprehensive profile delineating the evolution of cooperative acts remained an open task. Regarding this issue, spatial structure serves as a pivotal condition boosting the evolutionary advantage of cooperative behaviors~\cite{nowak2006five}. Such realistic constraints on social interactions of individuals allow the evolution of cooperation in the prisoner’s dilemma game~\cite{nowak1992evolutionary}, which, on the contrary, cannot evolve in the absence of an assortment between cooperators, like in a randomized or well-mixed population~\cite{nowak2004emergence}. The effects of network reciprocity have been explored mainly through numerical simulations~\cite{perc2017statistical,szolnoki2009resolving,rong2007roles,pan2024explaining,pi2025dynamic} and extended by some theoretical analysis~\cite{allen2017evolutionary,mcavoy2020social,su2022evolutionSA,meng2024dynamics,mcavoy2021fixation,zeng2025evolutionary}. These studies found that around $70\%$ of spatial structures can favor the evolution of cooperation and others, including complete graphs, favor the evolution of spite (an antisocial behavior)~\cite{allen2017evolutionary}. Therefore, how to {make cooperative behaviors evolve} within such structures remains simultaneously significant and has received substantial attention~\cite{fotouhi2018conjoining,su2022evolution}. It is noteworthy that beyond the classic pairwise dilemmas, the group dilemmas, i.e., multiplayer games, are also a crucial research direction in evolutionary games and significant advancements have been made in recent years~\cite{perc2013evolutionary, Wang2025emergence,sheng2024strategy,wang2024evolutionary,wang2026evolutionary}.

Traditional game theory often assumes that conditions characterizing the interactions of competitors are static and time-independent. This simplified assumption, however, cannot be maintained in several realistic situations. For example, environmental parameters (such as resource availability) inherently fluctuate over time due to different factors like migration, overexploitation of public goods, or climate change. Considering the influence of these temporal factors can reveal the coupling effects of human behaviors and environmental conditions~\cite{hilbe2018evolution}.  For instance, some recent studies have pointed out the profound impact of changing environments, which can lead to diverse evolutionary outcomes~\cite{stewart2014collapse,zeng2022spatial,weitz2016oscillating,feng2023evolutionary,perc2008social,tilman2020evolutionary,szolnoki2018environmental,assaf2013cooperation,ashcroft2014fixation}.

A recent work by Hilbe {\it et al.}~\cite{hilbe2018evolution} offered an intuitive and effective paradigm to analyze the coupling effect of individuals' behaviors and environment feedback in iterated games. The next step involved studying networked populations, and the corresponding critical threshold of the evolution of cooperation has been obtained in large homogeneous networks~\cite{su2019evolutionary}. Such a game transition (transitions among different games) pattern facilitates cooperative behaviors under birth-death and pairwise-comparison updating rules \cite{ohtsuki2006simple}, in which the evolution of cooperation is impossible on regular graphs~\cite{ohtsuki2006simple}. In some cases, the structure of social interactions in reality is finite and heterogeneous, which makes the mentioned threshold obtained on regular graphs~\cite{su2019evolutionary} inapplicable. Furthermore, the change of games is unnecessary to be fixed to game intensities. It is also reasonable to consider changing game types, driven by exogenous and endogenous changes~\cite{hilbe2018evolution,ashcroft2014fixation}. 

In this article, we provide a mathematical treatment to analyze the evolution of cooperative behaviors in a fixed social structure, represented by dyadic networks, where game rules can change over time, caused by the factors mentioned above. We present an analytical framework of the evolution of cooperative behavior in stochastic games, incorporating spatial heterogeneity and game diversity. Our analysis uncovers how distinct social dilemmas and population structures determine the effects of stochastic games in the presence and absence of mutation (random strategy exploration~\cite{traulsen2009exploration}). Our findings imply that stochastic games can exert conflicting influences on the evolution of cooperative behaviors in different social contexts instead of merely promoting cooperation. Finally, through simulations, we verify the phenomena obtained by theoretical analysis under weak selection and extend our results to strong selection. Therefore, our framework sheds light on whether and how stochastic games have a positive or negative effect on the evolution of cooperation.

\section{Results}
\begin{figure*}[ht]
    \centering
    \includegraphics[width=0.95\textwidth]{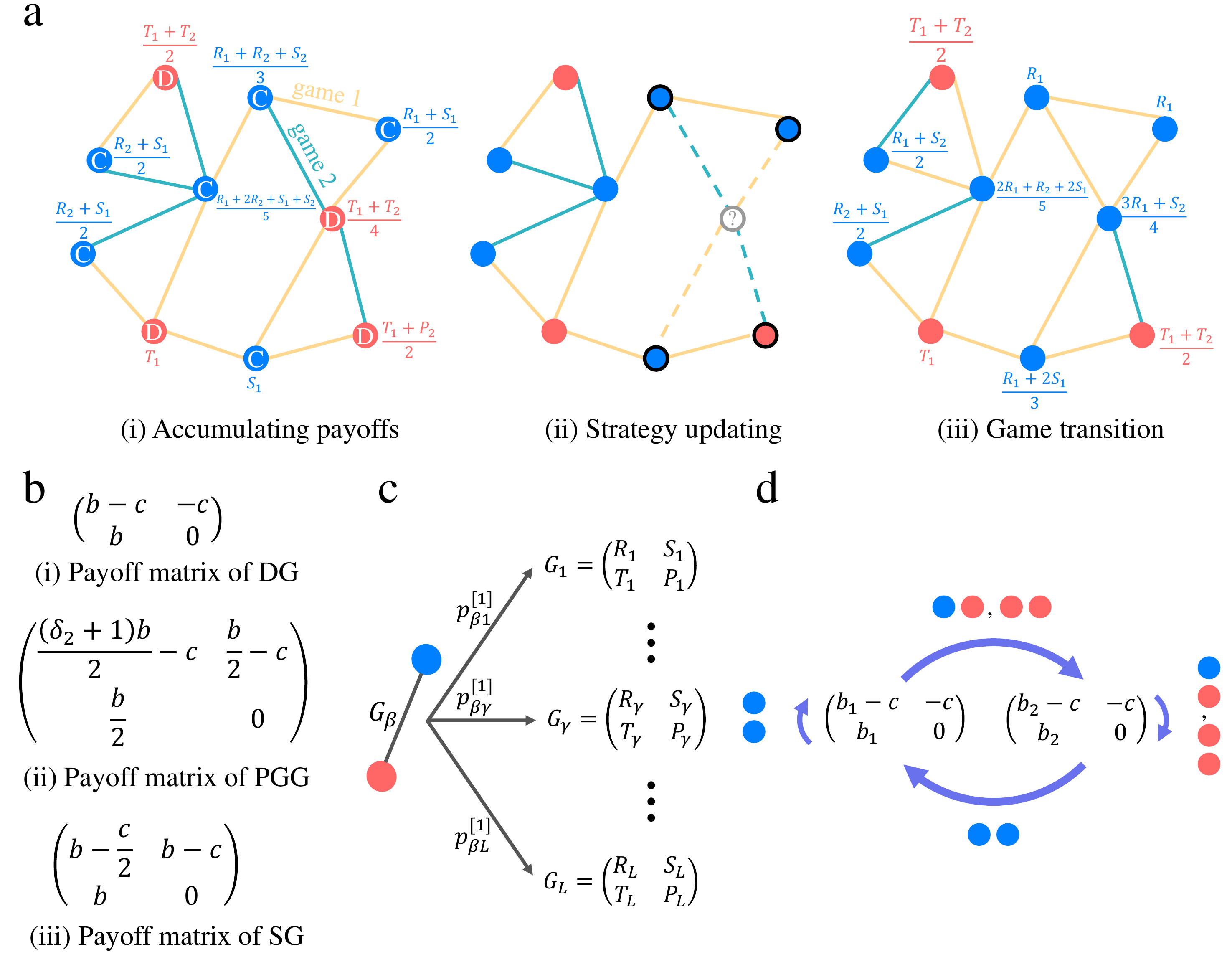}
    \caption{\textbf{Evolutionary dynamics in stochastic games on population structures.} {The population structure can be described by a graph, in which each node represents an individual and each edge represents the interaction between two individuals (panel~(a)). Each individual can choose one of the two strategies, Cooperation (blue node) and Defection (red node), in each round. The payoff of each individual is obtained by playing games (games are possibly different for different neighbors in the same round, illustrated by different edge colors, orange (cyan) edges representing game~$1$ ($2$)) with its neighbors, influenced by both its opponent's strategy and the current game (the payoff matrix of three considered games, i.e., the donation game (DG), the public goods game (PGG), and the snowdrift game (SG), are shown in panel (b)). After the accumulation of payoffs (averaged by the number of neighbors of each individual), an individual is uniformly chosen for death (gray circle) and all its neighbors (black circles) compete to reproduce at the empty site (dashed lines) proportional to their fitness, which is determined by their corresponding payoffs. Then, all games update, driven by either exogenous (independent of the population state) or endogenous factors (dependent on the population state) (panel~(c)). An example is given in panel~(d), in which mutual cooperation leads to a more profitable donation game (left game), and otherwise, a less profitable one would be reached (right game). The payoff structure of the donation game is a specific parameter setting of the general payoff matrix in panel~(c), i.e., $R=b-c$, $S=-c$, $T=b$, and $P=0$. The population eventually terminates in either an all-cooperator (fixation of cooperation) or all-defector (fixation of defection) state.}}
    \label{Fig1}
\end{figure*}

Evolution is considered on a structured population of $N$~nodes and $M$~edges. Its structure is described as an undirected and unweighted graph, in which nodes are individuals and edges are relationships between individuals (Fig.~\ref{Fig1}(a)). Each individual can adopt one of the following two strategies: Cooperation (C) or Defection (D). For contrast and better presentation of the function of stochastic games, we choose three typical social dilemmas for further analysis, i.e., the donation game (DG)~\cite{sigmund2010calculus}, the two-player public goods game (PGG)~\cite{hauert2006synergy,sheng2024strategy}, and the snowdrift game (SG)~\cite{hauert2004spatial,song2013coevolution}. In DG, cooperators pay a cost $c$ to produce a benefit $b$ for their co-player, and defectors pay no costs and generate no benefits (the ratio $b/c$ is known as the benefit-to-cost ratio, which plays an important role in further analysis). Particularly, for $b>c>0$, the donation game becomes a special instance of the prisoner’s dilemma game. In PGG, cooperators pay a cost $c$ to contribute toward the production of a public benefit, which is divided equally between players (no matter which strategies they took), and defectors pay no cost but benefit from the public goods. The biggest difference between DG and PGG is whether the benefits $b$ produced by cooperators are equally shared between the two players. Lastly, the feature of SG describes that players can enjoy the public benefit together without benefit allocation, and the cost is divided equally. Their corresponding payoff matrices are displayed in Fig.~\ref{Fig1}(b), respectively. These three representative examples cover major cooperation dilemmas inherent in nature. We denote the state of the whole population by $\mathbf{s}$ (length of $\mathbf{s}$ is $N$), where $s_i \in \{\textrm{C},\textrm{D}\}$ represents the strategy individual $i$ takes. 

After accumulating payoffs from interactions with neighbors, players update their strategies through imitation. In particular, an individual is randomly selected from the population for death, and all its neighbors compete to occupy the vacant place for birth in proportion to their fitness~\cite{ohtsuki2006simple}, known as the classic death-birth (DB) updating rule (see (ii) of Fig.~\ref{Fig1}(a)).

We introduce stochastic games to model the variance of environments. That is, the environment is reflected by the payoff matrix, described by four general parameters, and the payoff matrix is variable during evolution (see Fig.~\ref{Fig1}(c)). After each round of strategy updates, the games, taking placing between all connected two individuals, undergo an updating step. We consider all possible $L$ games during evolution, of which the state space is labeled as $E$. In each round, individuals find themselves in one of the $L$ possible states and will enter another game (possibly remain in the current game) in the next round. The probability of game $\beta$ transitioning to game $\gamma$ depends on the behaviors made by two individuals. In this case, there are possibly zero, one, and two cooperators, which can influence the quality of the environment, e.g., corresponding to the improvement and deterioration of the living environment. In the proposed model, this process is reflected by the changes in games, determined by individuals' strategies. Therefore, the stochastic process governing these game transitions can be described by three $L \times L$ matrices $\mathbf{P}^{[l]} = \left( p^{[l]}_{\beta \gamma} \right)$, in which $p^{[l]}_{\beta \gamma}$ is the probability of transition from game $\beta$ to game $\gamma$ given that there are $l \in \{ 0,1,2 \}$ cooperators in the considered pairwise interaction (e.g., there is a single cooperator in the given interaction in Fig.~\ref{Fig1}(c), i.e., $l=1$). Further technical details can be found in the corresponding sections in Supplementary Note~2. In the conventional and natural setting, mutual cooperation can provide a more profitable game. We denote $b_{\beta}$ ($b_{\gamma}$) as the benefit produced by cooperation in game $\beta$ (in game $\gamma$). If $b_{\beta} > b_{\gamma}$, we say that game $\beta$ is more profitable than game $\gamma$, assuming game $\beta$ and game $\gamma$ are the same game type. Without loss of generality, we set game $1$ as the most profitable game. We denote $\mathbf{g}$ (length of $\mathbf{g}$ is $M$) as the game state of the population, in which $g_i \in E$ means the game in the $i$-th interaction. 

To capture the scenario in which the game only has a small effect on strategy evolution, we set the selection strength $\delta$ to be weak ($0 < \delta \ll 1$)~\cite{wu2010universality}. The fitness of each individual is measured by $F_i(\mathbf{s}, \mathbf{g}) = 1 + \delta f_i(\mathbf{s}, \mathbf{g}) + O(\delta^2)$, where $f_i(\mathbf{s}, \mathbf{g})$ represents the averaged payoff of individual $i$ under population state vector $\mathbf{s}$ and {game state} vector $\mathbf{g}$.

Changes in environments can be caused by a complex interplay between intrinsic and extrinsic factors. For simplicity and better elaboration, we focus on transitions driven by either exogenous or endogenous influences. Furthermore, we consider the simple setup in the main text, in which the game types are out of DG, SG, and PGG, and two possible games (game types can be different) are considered (i.e., $L=2$). An illustration of our model is depicted in Fig.~\ref{Fig1}. 

\subsection{General Rule for Evolution of Cooperation}
Only one strategy, either cooperation or defection, will survive after a sufficiently long time of evolution without mutation. Let $\rho_{\textrm{C}}$ ($\rho_{\textrm{D}}$) denote the probability that a randomly initialized cooperator (defector) successfully propagates its strategy to the entire population consisting of all defectors (cooperators) otherwise. Our primary focus lies in under which condition $\rho_{\textrm{C}}$ is greater than $1/N$, known as the absolute measure of success, quantifying the outperforming of cooperation over neutral evolution, under which the fixation probability is $1/N$. We also carry out analysis on the relative measure of success $\rho_{\textrm{C}}>\rho_{\textrm{D}}$, quantifying the outperforming of cooperation over defection.

A general pairwise game can be described by four parameters: $R$ (Rewards for mutual cooperation), $S$ (Sucker's payoff for unilateral cooperation), $T$ (Temptation for unilateral defection), and $P$ (Punishment for mutual defection). We have proven that cooperation is favored in the sense of $\rho_\textrm{C} > 1/N$ on a given population structure $\mathcal{G}$, when 
\begin{equation}
\substack{ \left( S - R \right) \sum\limits_{i,j \in \mathcal{G}} d_i p^{(1)}_{ij} \eta_{ij} \\+\\ \left( S - P \right) \sum\limits_{i,j \in \mathcal{G}} d_i p^{(2)}_{ij} \eta_{ij} \\+\\ \left( T - P \right) \sum\limits_{i,j \in \mathcal{G}} d_i p^{(3)}_{ij} \eta_{ij}} > \substack{ (S + T - R - P)} \sum_{i,j,k \in \mathcal{G}} d_i p^{(2)}_{ij}p^{(1)}_{jk} \eta_{ijk}\,,
\label{eq1}
\end{equation}
in which $d_i$ is the number of neighbors of individual $i$ (also degree of $i$), $p^{(1)}_{ij}$ ($p^{(n)}_{ij}$) represents one-step ($n$-steps) probability of random walk from $i$ to $j$, $\eta_{ij}$ ($\eta_{ijk}$) means the expected time for two (three) random walks starting from nodes $i$ and $j$ ($i $, $j$, and $k$) to meet at a common node (i.e., the coalescence time)~\cite{allen2017evolutionary}. For simplicity, we take the following notations $\eta_{(n)} = \sum_{i,j \in \mathcal{G}} \pi_i p^{(n)}_{ij} \eta_{ij}$ and $\Lambda_{(n)} = \sum_{i,j,k \in \mathcal{G}} \pi_i p^{(n)}_{ij}p^{(1)}_{jk} \eta_{ijk}$, where $\pi_{i} = d_i / \sum_{j \in \mathcal{G}}d_j$ is the weighted degree of $i$. Intuitively, $\eta_{(n)}$ ($\Lambda_{(n)}$) quantifies the assortment of strategies between two nodes $n$ steps away (resp. among three nodes, in which the first two nodes are $n$ steps away, and the last two nodes are neighbors). These quantities can be calculated directly once the population structure is given.

We can calculate stochastic games' corresponding critical benefit-to-cost ratio based on Eq.~\ref{eq1} (see Methods). When it comes to specific games, e.g., the aforementioned donation game and snowdrift game, we are concerned with the critical benefit-to-cost ratio $(b/c)^*$ for $\rho_{\textrm{C}} = 1/N$. We say that cooperation is favored when $b/c>(b/c)^*$, provided that $(b/c)^*$ is positive. If $(b/c)^*$ is negative, spite is favored~\cite{allen2017evolutionary}. As a starting point, we first consider exogenous game transitions.

\subsection{Exogenous Game Transitions}

\begin{figure*}[!ht]
    \centering
    \includegraphics[width=0.98\textwidth]{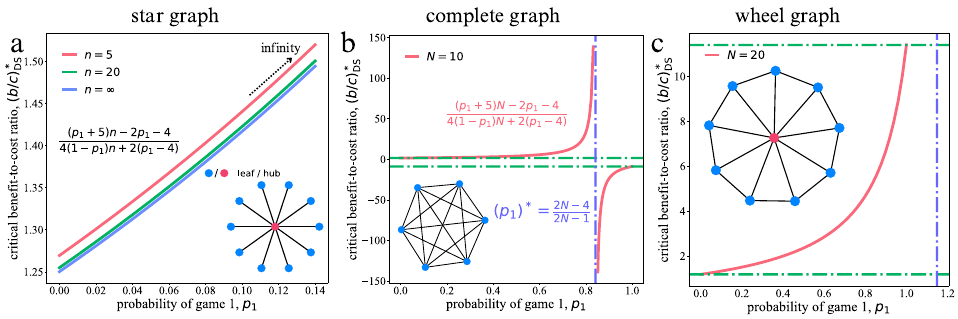}
    \caption{{\textbf{Evolutionary game with exogenous game transitions on diverse population structures under death-birth updating.} We present the critical benefit-to-cost ratio, $(b/c)^*_{\text{DS}}$, as a function of the probability of the donation game (game 1), $p_1$, on the star graph with one central node and $n$ peripheral nodes (panel~(a)), complete graph with size $N$ (panel~(b)), and wheel graph with size $N$ (panel~(c)). These graphs represent structures in which the critical ratio is infinite, negative, and positive, respectively. Panels show different trajectories of $(b/c)^*$ as $p_1$ changes. The vertical line is the discontinuous point $(p_1)^*$, which leads to $(b/c)^* = \infty$, and the horizontal line indicates the critical benefit-to-cost ratio in a single game, i.e, $p_1 = 0$ (corresponding to the single snowdrift game) and $p_1 = 1$ (corresponding to the single donation game). In general, $(p_1)^*$ is within the range $[0,1]$ when the critical ratio in the donation game is negative. In contrast, the positive critical ratio leads to $(p_1)^*$ out of $[0,1]$, the discontinuous change of $(b/c)^*$ (vertical line in panel~(c)), therefore, cannot be observed. The insets of each subgraph are the corresponding network structures.}} 
    \label{Fig2}
\end{figure*}

Exogenous game transitions refer to the stochastic transition of games that is independent of the evolutionary process. For simplicity, we assume that the environment enters game $\beta$ with probability $p_{\beta}$ in each round, in which cooperation brings $b_{\beta}$ benefits. Analytical results are available in the Supplementary Note~3.4. 

We can see that the effect of such a transition mode is the same for different social dilemmas. Moreover, if transitions occur among profitable games more frequently, the critical ratio is lower, and vice versa. Thus, the best choice is to always stay in the most profitable game. Naturally, such a transition mechanism can promote (inhibit) cooperation only if transitions are within more (less) profitable games. Specifically, when $p_1 = 1$ (i.e., evolution with a single game), we recover the critical benefit-to-cost ratio for each corresponding game. 

Our analysis above assumes that the game transitions are homogeneous, where transitions are among single cooperative games, e.g., confined to only DG or SG. However, considering the heterogeneity of games~\cite{hilbe2018evolution,su2019edgediversity}, for simplicity, we suppose that DG and SG are uniformly switched between each other during evolution, with probabilities $p_1$ and $p_2$ ($L=2$ and $p_1 + p_2 = 1$), respectively. The possible value of the critical benefit-to-cost ratio in the donation game can be classified as infinite value, negative value, and positive value (the star graph, complete graph, and wheel graph as typical examples, respectively, and the structures of each graph are displayed in Fig.~\ref{Fig2}). We take the three representative structures mentioned as examples to explicitly exhibit all possible changing trajectories of each kind of critical benefit-to-cost ratio with $p_1$. 

We begin with star graphs (composed of $n$ lead nodes and $1$ hub node). Recall that the critical benefit-to-cost ratio for star graphs is infinite in the donation game, i.e., $(b/c)^* = \infty$. In other words, cooperation can never evolve in star graphs. A possible solution to avoid the infinite critical ratio is to join another (star) graph to make the ratio finite~\cite{allen2017evolutionary,fotouhi2018conjoining}. Indeed, without conjunction, the introduction of some other social dilemmas, like SG, thus breaking the singularity of the game, can facilitate the evolution of cooperation in star graphs. The introduction of SG transfers the infinite critical ratio to a finite value, which is determined by both $n$ and $p_1$ (Fig.~\ref{Fig2}(a)). The intuition for this result is straightforward: the introduction of SG neutralizes the hub’s structural dominance on star graphs. In the pure DG setting ($p_2=0$), the central defector has an absolute advantage. While the introduction of SG imposes a risk of ``punishment'', due to its low inherent payoff setting for mutual defection (\textrm{D}-\textrm{D} is the worst strategy pair in SG, while this is not the case in DG). It weakens the hub’s structural dominance, thus ensuring a finite $(b/c)^*$.

Our result shows that cooperation can be rescued for any given $p_1 < 1$, since the corresponding critical ratio changes from an infinite value to a finite one. The explicit change of $(b/c)^*_{\text{DS}}$ with respect to $p_1$ is presented in Fig.~\ref{Fig2}(a).

Next, we take complete networks into account. It is well-recognized that spite evolves in it, with size $N$, due to its negative critical ratio in DG, yielding $(b/c)^* = -N+1$. However, the introduction of the aforementioned transition pattern can allow the evolution of cooperation under some conditions since the corresponding critical ratio is a positive value under certain $p_1$ (the left red curve in Fig.~\ref{Fig2}(b)).

Furthermore, we show that there always exists an unusual value $(p_1)^*=(2N-4)/(2N-1)$ making the critical ratio infinite (vertical line in Fig.~\ref{Fig2}(b)). Therefore, the improper value of $p_1$ ($p_1$ around $(p_1)^*$) can incur an unaffordable benefit-to-cost ratio to support the evolution of cooperation. Furthermore, the corresponding critical benefit-to-cost ratio is positive for $p_1 < (p_1)^*$ and becomes negative but smaller than the critical ratio in DG for $p_1 > (p_1)^*$ (red curve in Fig.~\ref{Fig2}(b)). Based on this, we conclude that when $p_1 < (p_1)^*$, cooperation evolves in complete networks and spite is inhibited, when $p_1 > (p_1)^*$. 

Finally, we present the results obtained on wheel graphs (Fig.~\ref{Fig2}(c)). Specifically, as shown in Fig.~S1(b), for the small wheel graph ($N \leq 8$), its corresponding critical benefit-to-cost ratio in DG is negative, and it becomes positive otherwise ($N > 8$). Furthermore, when $N$ is small, a similar trend can be observed (see Fig.~S1(c) and Fig.~\ref{Fig2}(b) for comparison); for greater size, the critical ratio exhibits smooth changes without breaking in Fig.~\ref{Fig2}(c), from the critical ratio in the single SG (bottom horizontal line) to the critical ratio in the single DG (top horizontal line). Therefore, when accounting for such game heterogeneity, its critical benefit-to-cost ratio falls between the respective thresholds identified in SG and DG. 

Taken together, we have analyzed how such game heterogeneity affects the evolutionary outcomes in all cases, including the infinite critical ratio (Fig.~\ref{Fig2}(a)), the negative critical ratio (Fig.~\ref{Fig2}(b)), and the positive critical ratio (Fig.~\ref{Fig2}(c)) in DG. Other structures present analogous trends of change (belonging to one of the trajectories in Fig.~\ref{Fig2}), once their corresponding critical benefit-to-cost ratios in the donation game are known. This mechanism not only enables the evolution of cooperation on structures in which cooperation cannot evolve (e.g., the star graph and complete graph) but also reduces the barrier to the success of cooperation (e.g., the large wheel graph). 

Related numerical simulations with the exogenous game transition on random regular graphs are shown in Fig.~S2. In the following part, we discuss the endogenous game transition.

\subsection{Endogenous Game Transitions}
Endogenous game transitions refer to the stochastic transitions of games that depend on the evolutionary process. That is, every update of games is uniquely governed by the strategy pair adopted by the two players. For that purpose, we focus on one classic game transition pattern, i.e., the deterministic state-independent game transition~\cite{hilbe2018evolution,su2019evolutionary}. In detail, mutual cooperation (i.e., $l=2$) leads to a more profitable game (game 1), and other strategy pairs (unilateral cooperation and mutual defection, i.e., $l=1$ and $l=0$) result in a less profitable game (game 2). Fig.~\ref{Fig1}(d) illustrates the transition process (DG as an example), in which $b_1$ is greater than $b_2$, indicating the increase in profit caused by the transition from game $2$ to game $1$, as a reward for mutual cooperation. Therefore, the process can be regarded as the dynamic adjustment of environments, i.e., the improvement of the environment for mutual cooperation, and its destruction otherwise. 

\begin{figure*}[!ht]
    \centering
    \includegraphics[width=0.95\textwidth]{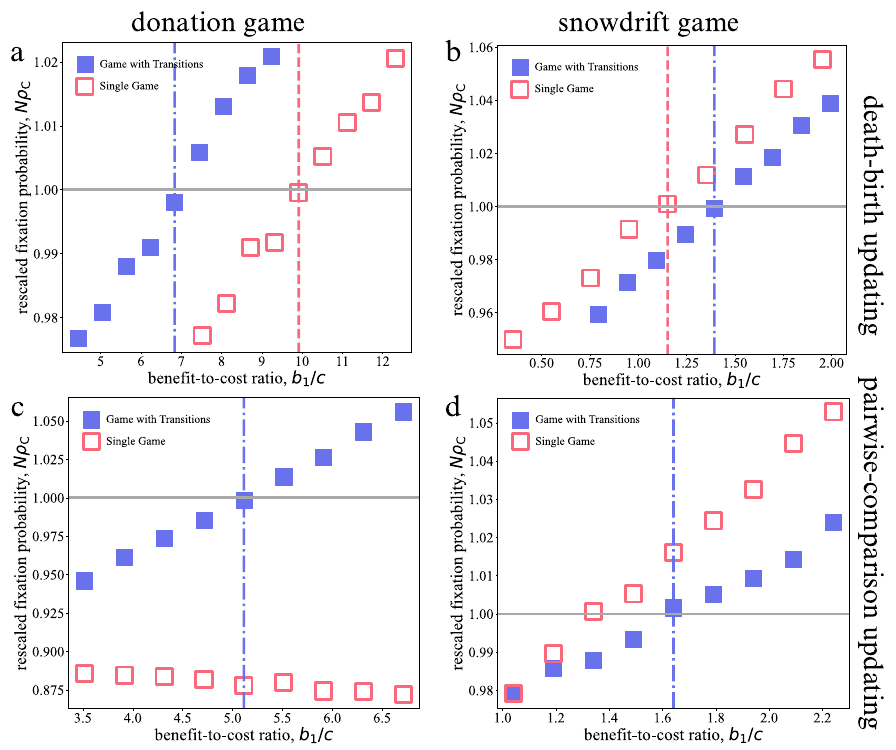}
    \caption{\textbf{Evolutionary dynamics with endogenous game transitions.} We consider the death-birth (DB) (panels~(a) and (b)) and pairwise-comparison (PC) updating (panels~(c) and (d)) in the donation game (DG) (panels~(a) and (c)) and snowdrift game (SG) (panels~(b) and (d)). The fixation probability of cooperation, $\rho_{\textrm{C}}$, is presented as a function of the benefit-to-cost ratio $(b_1/c)$ on Barab\'{a}si-Albert (BA) graph~\cite{barabasi1999emergence}. Cooperation is favored if $N\rho_\textrm{C}$ exceeds $1$ (horizontal lines), i.e., $\rho_{\textrm{C}}>1/N$. Filled (open) squares are the results of evolution with game transitions (resp. single game) obtained from Monte Carlo simulations. Vertical lines represent corresponding analytical predictions of the critical benefit-to-cost ratios $(b_1/c)^*$, above which the evolution of cooperation is facilitated. All games are initialized as game $2$. Simulations on fixation probabilities are repeated for $10^7$ independent runs. The parameters under DB are: the benefit difference between two games $\Delta b = 1$ in DG and $\Delta b = 0.6$ for SG. The parameters under PC are: the benefit in game 2 $b_2 = 2$ in DG and $b_2 = 1$ in SG. Other parameters are: BA with linking number $m=3$, the cost for cooperation $c = 1$, and selection strength $\delta = 0.01$.}
    \label{Fig3}
\end{figure*}

We derive the corresponding critical benefit-to-cost ratio applied to arbitrary population structures (see Methods). Transitions between two games can effectively promote cooperation in DG (Fig.~\ref{Fig3}(a)) and inhibit cooperation in SG (Fig.~\ref{Fig3}(b)). For simplicity, we fix the cost $c=1$ as the default. For example, setting $b_1 = 7$ (a value larger than $6.84$ corresponding to the blue vertical line in Fig.~\ref{Fig3}(a)) and $b_2 = 6$  in DG can favor the evolution of cooperation, though, in every single game, the evolution is disfavored since the required benefit is about $9.92$ (red vertical line in Fig.~\ref{Fig3}(a)). In contrast, even though the benefit in game 1, $b_1 > 1.15$ (the required benefit in the single SG, corresponding to the red vertical line in Fig.~\ref{Fig3}(b)). In this sense, cooperation can evolve without game transitions in SG. The intuition for the phenomenon is: in SG, the player's best strategy is the opposite of their opponent's, determined by the payoff setting; however, the game switching can strengthen the clusters of cooperators in DG. Therefore, though the switch of games favors the evolution of cooperation in DG, it disfavors cooperation in SG.

We also explore whether such a game transition can accelerate/decelerate the evolutionary process. Our simulations (Fig.~S3) show that it has an unnoticeable influence on evolution time, both the time for cooperation to fix (fixation time) and the time for absorption (absorbing time). Therefore, this transition pattern can effectively reduce the requirement for the fixation of cooperation in DG, but it cannot accelerate the fixation process.

Next, we consider the pairwise-comparison updating, under which a randomly chosen individual (denoted $i$) chooses one of its neighbors at random (denoted $j$) and imitates its strategy with probability $\left(1+\exp\left\{-\delta\left(f_{j}-f_{i}\right)\right\}\right)^{-1}$~\cite{szabo1998evolutionary}. In this update, the game transition can favor the evolution of cooperation in DG, while it cannot in a single DG (Fig.~\ref{Fig3}(c)). Evolutionary outcomes are similar to those under death-birth updating in the SG (Fig.~\ref{Fig3}(d)). We refer the reader to Supplementary Note~7 for technical details.

To explore the effectiveness of our methods, we also test a representative heterogeneous population structure, which is the Barab\'{a}si-Albert (BA) scale-free graph~\cite{barabasi1999emergence}. Our simulations, shown in Fig.~\ref{Fig3}, exhibit a high degree of concordance with the theoretical predictions derived from our analytical framework.

\subsection{Complete Networks}
We take some specific structures for further investigations under the death-birth updating. But first, we consider a complete network of size $N$, where each node is connected to all other nodes. It is well-known that spite, instead of cooperation, evolves in it since its negative critical benefit-to-cost ratio $(b/c)^{*} = -N + 1$ in DG~\cite{allen2017evolutionary}. 

We have the exact expressions for DG and SG under the deterministic state-independent game transition (see Supplementary Note~3 for details), that is
\begin{equation}
\left( \frac{b_1}{c}\right)^{*}_{\text{DG}} = -N + 1 + \frac{N+1}{3} \cdot \frac{\Delta b}{c}\,,
\label{W-M DG}
\end{equation}
and
\begin{equation}
\left( \frac{b_1}{c} \right)^{*}_{\text{SG}} = \frac{5N-4}{4\left( N - 2 \right)} + \frac{\Delta b}{2c}\,,
\label{W-M PGG}
\end{equation}
where $\Delta b = b_1 - b_2$ represents the benefit difference in two games.

Intriguingly, according to Eq.~\ref{W-M DG}, $\left( b_1/c \right)^{*}_{\text{DG}} = 2$ is a constant if $\Delta b / c = 3$, indicating that the critical ratio is independent to $N$. Furthermore, the success of the evolution of cooperation requires $\Delta b / c > 3$ in complete networks, which turns the negative threshold into a positive one. Therefore, such changes in games are another effective approach, from a different perspective of exogenous transition patterns, to rescue cooperative behaviors under certain mild conditions. Compared to DG, Eq.~\ref{W-M PGG} implies that the size $N$ has no impact on game transitions in SG. Moreover, $\left( b_1 / c \right)^{*}_{\text{DG}}$ and $\left( b_1 / c \right)^{*}_{\text{SG}}$ both escalate in response to an increment in $N$.  However, the former's growth is unbounded, while the latter converges to a finite value.

\subsection{Ceiling Fan Networks and Conjoined Star Networks}
\begin{figure*}[!ht]
    \centering
    \includegraphics[width=0.90\textwidth]{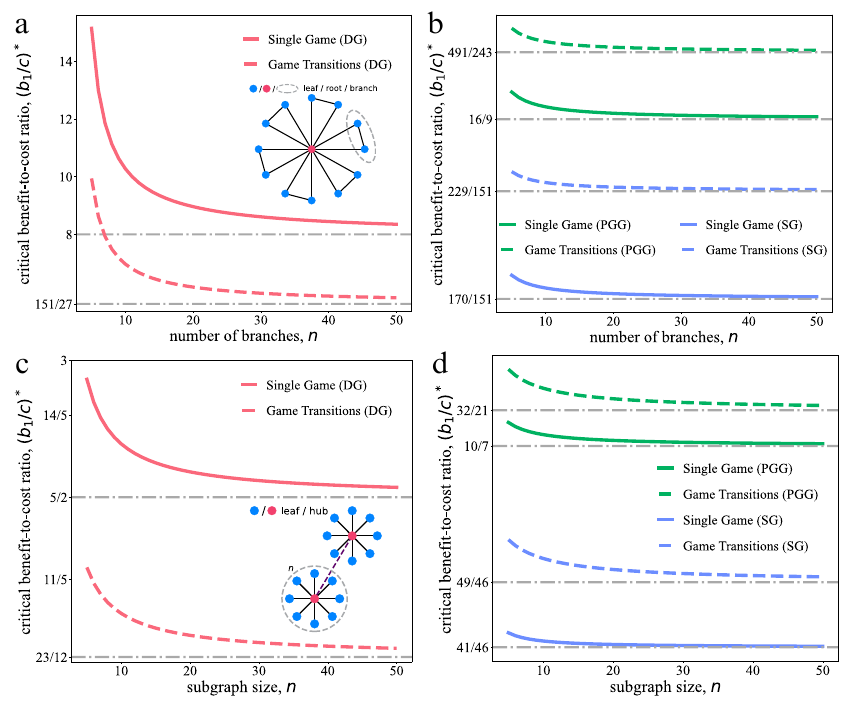}
    \caption{\textbf{Effects of game transitions on ceiling fan graphs and conjoined star graphs.} The critical benefit-to-cost ratio $(b_1/c)^*$ for cooperation as a function of the number of branches $n$ on ceiling fan graphs (panels~(a) and (b)) and the subgraph with size $n$ on conjoined star graphs (panels~(c) and (d)). The dashed (resp. solid) curves correspond to the analytical results for finite $n$ varying from $5$ to $50$ under deterministic game transitions (resp. single game). The dash-dotted horizontal lines display theoretical results for infinite $n$. The three cooperative games are the donation game (DG), public goods game (PGG), and snowdrift game (SG), as indicated in the legend.}
    \label{Fig4}
\end{figure*}

Next, we use ceiling fan networks and conjoined star networks as two non-trivial and symmetric topological structures for further analysis on the effects of population size. In detail, each of the $n$ leaves of a star is connected by an edge to one other in ceiling fan networks (Fig.~\ref{Fig4}(a)) and conjoining two identical star graphs with size $n$ via hubs generates conjoined star networks (Fig.~\ref{Fig4}(c)). Then, due to the symmetry of these graphs, we can obtain the exact expressions for each structure, according to our analytical framework (see Supplementary Notes~3 and 4 for a detailed derivation and exact expressions), and the variances of $(b/c)^*$ with $n$ in the context of DG, PGG, and SG are displayed in Fig.~\ref{Fig4}.

In the limit of many branches (resp. subgraph size), $n \to \infty$, the deterministic transition can reduce $30.1\%$ (resp. $23.3\%$) value of the critical benefit-to-cost ratio in DG, from $8.0$ to $5.6$ (resp. from $2.5$ to $1.92$), under $\Delta b/c = 1$. However, it can increase the value of $34.7\%$ (resp. $19.5\%$) of the critical ratio in SG, from $1.13$ to $1.52$ (resp. from $0.89$ to $1.07$), in the same setting. The exact values for each case are given in Fig.~\ref{Fig4}.

In DG, the deterministic state-independent game transition reduces the threshold for the {evolution of cooperation}, while it is elevated in the context of PGG and SG (as demonstrated in Fig.~\ref{Fig3} and Fig.~\ref{Fig4}). In other words, the effect of such a game transition is beneficial to cooperative behavior in DG but harmful to it in PGG and SG. The difference originates from the fact that the coefficient of the term describing the effects of the game transition is negative in DG, but the corresponding coefficients in both PGG and SG are positive instead. It indicates that whether such a deterministic state-independent game transition can promote cooperative behavior is determined by both the topological and payoff structures. 

\subsection{Results for Empirical Social Networks}

\begin{figure*}[!ht]
    \centering
    \includegraphics[width=0.95\textwidth]{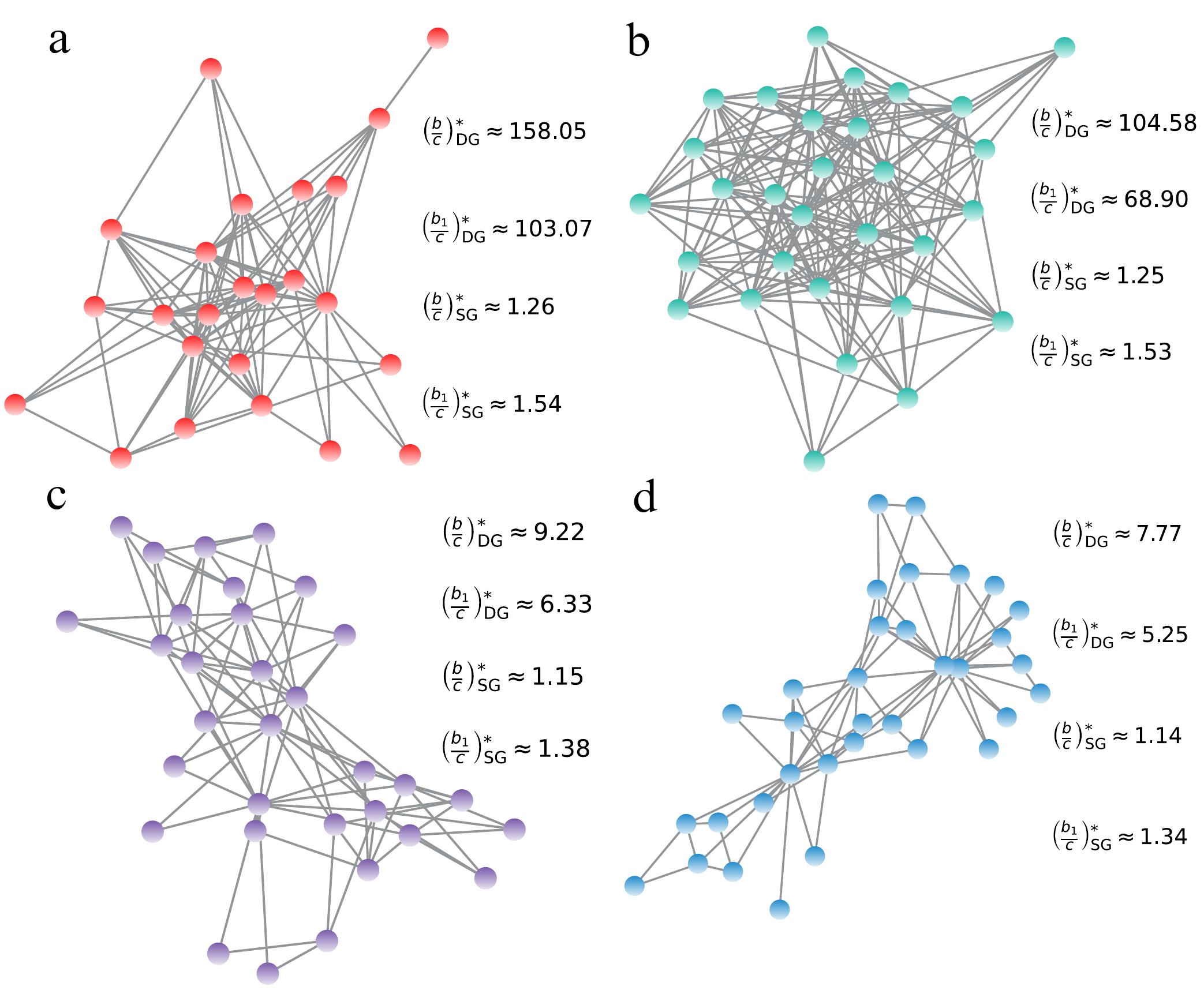}
    \caption{\textbf{Effects of game transitions on empirical social networks {under death-birth updating}.} Four empirical datasets are taken into account: (a) grooming {behaviors} observed among a cohort of 24 primates~\cite{griffin2012community}; (b) social associations of a group comprising 30 beetles~\cite{formica2016consistency}; (c) the friendship network of its $34$ members within a karate club~\cite{zachary1977information}; (d) the social connections among a sample of 32 women from the Davis community~\cite{davis1941deep}. The theoretical predictions under death-birth updating are provided for each social network in the context of the donation game $(b/c)^*_{\text{DG}}$ (resp. snowdrift game $(b/c)^*_{\text{SG}}$) and its corresponding value under the deterministic game transition $(b_1/c)^*_{\text{DG}}$ (resp. $(b_1/c)^*_{\text{SG}}$).}
    \label{Fig5}
\end{figure*}

To demonstrate the broad validity of our observations, we also analyze five empirical social networks derived from diverse social contexts (Fig.~\ref{Fig5} and Fig.~S6): an animal social network of $24$ primates based on observation of grooming interaction~\cite{griffin2012community}; an animal social network of $30$ beetles based on spatial proximity, where the social partners are defined as any beetle within $3$ cm (i.e., approximately $2$ body lengths) of the focal beetle~\cite{formica2016consistency}; a social network containing social ties among the members of a university karate club collected by Wayne Zachary in 1977~\cite{zachary1977information}; a social network recording interactions between $32$ Davis Southern women from 1930s~\cite{davis1941deep}; a social network describing frequent associations in dolphins~\cite{lusseau2003bottlenose}. Network statistics are displayed in Tab.~\ref{tab:1}, where $|V|$, $|E|$, $k_{\mathrm{avg}}$, $k_{\mathrm{max}}$ and $D$ represent the number of vertices, the number of edges, the average degree, the maximum degree, and the network density, respectively.
\begin{table}[!htb]
    \centering
    \caption{Network Statistics. The key features of the considered networks are listed in the table below.}
    \begin{tabular}{c|c|c|c|c|c} \hline
    
        Population &  $|V|$ & $|E|$ & $k_{\mathrm{avg}}$ & $k_{\mathrm{max}}$ & $D$\\ \hline
        {Macaca fuscata} & $24$ & $103$ & $8.6$ & $16$ & $0.37$ \\ \hline
        {Bolitotherus cornutus} & $30$ & $185$ & $12.3$ & $19$ & $0.43$ \\ \hline
        Karate club &  $34$ & $78$ &  $4.6$ & $17$ & $0.14$ \\ \hline
        Davis Southern women &  $32$ & $89$ &  $5.6$ & $14$ & $0.18$ \\ \hline
        {Bottlenose dolphin} & {$62$} & {$159$} &  {$5.1$} & {$12$} & {$0.08$} \\ \hline
    \end{tabular}
    \label{tab:1}
\end{table}

We calculate the critical ratios for {both DG and SG}, as two representative dilemmas, $(b/c)^*_{\text{DG}}$ and $(b/c)^*_{\text{SG}}$ for each empirical structure (Fig.~\ref{Fig5}). With the introduction of game transitions, cooperation evolves harder (easier) in the snowdrift (donation) game since a higher (lower) benefits-to-cost ratio $b/c$ is required. The results of real-world networks are consistent with those obtained on synthetic networks.

\subsection{Effects of Game Transitions on Costs}
We have discussed the case where the game transition pattern only dynamically adjusts the benefits according to strategy pairs, leading to different outcomes in diverse social dilemmas. From another perspective, it is reasonable to analyze how dynamic games regulate the costs of cooperation, affecting evolutionary outcomes. In detail, when mutual cooperation is achieved, the cooperators pay less $c_1$ as the cost of altruistic behaviors and pay a higher cost $c_2$ ($c_2 > c_1$) otherwise. Different cost values generate the same benefit $b$ for its opponent. Since an increase in cost is harmful to the evolution of cooperation, we define a critical cost $c^*$. For $0<c<c^*$ (resp. $c>c^*$), we can say that the evolution of cooperation is favored (resp. disfavored). 

Our results show that the presence of such a transition pattern can reduce the value of $c^*$ both in DG (Fig.~S4(a)) and SG (Fig.~S4(b)) under death-birth updating, which is harmful to the evolution of cooperation.

\subsection{Relative Measure of Success}
We have analyzed the critical ratio for the dominance of cooperation over the neutral fixation probability $1/N$. It is also reasonable to explore the results for the threshold of cooperation outperforming defection in evolution, denoted by $\rho_{\mathrm{C}} > \rho_{\textrm{D}}$. We focus on the homogeneous structure, i.e., the random regular graph, described by two parameters, its size $N$ and degree $k$ (i.e., all nodes have the same number of $k$ neighbors). For technical details, we refer the reader to Supplementary Notes~5 and 8.

Recall that under the assumption that $N$ is much larger than $k$, the condition of $\rho_{\mathrm{C}} > \rho_{\textrm{D}}$ is $\Delta b/c>2$ under pairwise-comparison updating in DG~\cite{su2019evolutionary}, where $\Delta b = b_1 - b_2$ and $b_1$ (resp. $b_2$) is the benefit in game $1$ (resp. game $2$). It indicates that the exact values of $b_1$ and $b_2$ have no impact on the success of the evolution of cooperation against defection; in contrast, only the difference between $b_1$ and $b_2$ does. However, we find that the relation between $b_2$ and $N$ is subtle under pairwise-comparison updating. Fig.~S5 exhibits how the value of $b_2$ affects the critical ratio $(\Delta b/c)^*$. As we can see, for small $b_2$ (e.g., $b_2 = 2$ and $b_2 = 3$), the critical ratio $(\Delta b/c)^*$ can be approximated as $2$, which is exactly the known result that the difference between benefits matters. However, when $b_2$ equals $N$ ($b_2 = N = 100$ in this case), $(\Delta b/c)^*$ is around $4$, which greatly deviates from the expected result that $(\Delta b/c)^*$ is $2$ (green curve in Fig.~S5). Furthermore, the condition of whether $b_2$ makes a difference in evolutionary outcomes lies in whether $N$ is simultaneously much larger than $k$ and $b_2$. Therefore, we stress that the exact value of $b_2$ can also be a decisive factor if the ratio $b_2/N$ is non-negligible.

In general, in the context of DG and relative measure of success under death-birth updating, the deterministic game transition always reduces the critical threshold of facilitation of cooperation (Fig.~S6(b)). The same setting, however, has no impact on evolutionary outcomes in SG (Fig.~S6(c)).

\section{Discussion}
In this study, we have assumed that the change of environments affects the benefit/cost produced by a cooperative behavior, which can be regarded as the synergistic benefit for mutual cooperation and is intrinsically linked to partner-fidelity feedback mechanisms~\cite{foster2006general,sachs2004evolution}. Partner-fidelity feedback refers to the mechanism by which the fitness of a cooperator's partner can be promoted by a cooperating act, which also enhances the fitness of the cooperator itself in return. Otherwise, the failure of cooperation ultimately leads to the reduction of the individual's own fitness.

The payoff structure can be described as static (e.g., Eqs.~\ref{effective DG} and \ref{effective SG}), though the games are dynamic during the evolution due to their corresponding deterministic transition pattern (see Methods for more details). From another perspective, our results show that reducing the reward, $b_1$ in game 1, for mutual cooperation (i.e., $b_1 < b_2$) can favor the evolution of cooperation in the snowdrift game. In contrast, it undoubtedly hinders the evolution of cooperation in the donation game. The way to favor the evolution of cooperation is to simultaneously increase mutual and unilateral cooperation benefits in the snowdrift game. Furthermore, investigations on reducing the cost of altruistic behavior as a reward for mutual cooperation show that it narrows the survival space of cooperation. This finding holds for both the donation game and the snowdrift game. Our systematic analysis uncovers the diverse impacts of dynamic games, driven by endogenous or exogenous factors, on evolutionary outcomes in different social dilemmas. Our results can be useful for giving more accurate predictions and providing more adequate reactions when facing varying environments and different imitation rules to enable the sustainable propagation of altruistic behaviors in social systems. As an application for collective cooperation among agents, they offer significant benefits for designing and optimizing game rules, e.g., whether it is more profitable to increase the benefits or reduce costs as a reward to allow the emergency and maintenance of group intelligence in autonomous systems.

We have revealed that game transitions allow the evolution of cooperation in the donation game, but can also lead to its inhibition in the public goods game and snowdrift game. The result for the donation game is in line with~\cite{hilbe2018evolution} under intermediate (neither weak nor strong) selection, where game transitions favor the evolution of cooperative behaviors. Meanwhile, the result for the public goods game breaks the conventional understanding of stochastic games under strong selection, which can induce a substantial increase in cooperation levels~\cite{hilbe2018evolution}. Therefore, to the best of our knowledge, our findings could be helpful to provide a novel insight into how to resolve the tragedy of the commons~\cite{hardin1968tragedy}, from the perspective of how to adjust benefits/costs as the rewards/punishments for cooperation facing different social dilemmas.

As pointed out in~\cite{wu2013extrapolating}, results that hold for weak selection cannot be extended to stronger selection, and the ranking of strategies can vary with selection strength for multiplayer games. As we have emphasized, our research is carried out under the assumption of weak selection. A recent study suggests that there is no efficient algorithm for arbitrary selection intensity~\cite{jensen2015computational}, and some limited results are obtained for some highly symmetrical structures~\cite{altrock2017evolutionary}. Therefore, quantitative analysis would become notoriously complicated for stronger selection intensities and heterogeneous structured populations. Through simulations, we find that the effect of game transitions in each social dilemma persists in strong selection (DG in Fig.~S7(a)-(c), PGG in Fig.~S7(d)-(f), and SG in Fig.~S7(g)-(i)). Furthermore, in the donation game, some population structures can support the dominance of cooperation only with game transitions (Fig.~S7(b)), and neither of them can be supported in some spatial structures (Fig.~S7(c)).  

{In general,} heterogeneity plays a vital role in evolutionary dynamics, as explored in spatial structures~\cite{santos2005scale,allen2015molecular,zhou2021aspiration}, payoffs~\cite{amaral2016evolutionary,wang2014different}, and games~\cite{hilbe2018evolution,su2019edgediversity,feng2023evolutionary}. Based on the heterogeneous game transition, we provide analytical predictions of it. As we argued, cooperation can be promoted in star graphs, beneficial from exogenous transitions with heterogeneity. It implies that the introduction of another game helps to resolve the dilemma faced in a single game. Moreover, a human experiment suggests that the evolutionary outcomes do not solely depend on the nature of the social dilemma but are also significantly influenced by the class of individuals (e.g., gender) participating~\cite{kummerli2007human}.

Our results for fixation apply to mutation-free evolutionary dynamics since if mutation or random strategy exploration~\cite{traulsen2009exploration} is allowed, evolution would eventually enter one ``mutation-selection'' stationary distribution~\cite{allen2014measures,allen2019mathematical}. A recent work proposes a powerful framework to analyze this process theoretically~\cite{mcavoy2022evaluating}. Based on this, we can directly calculate the critical ratio for the mean cooperation frequency $\left<f_{\textrm{C}}\right>$ greater than $1/2$ (i.e., the mean cooperation frequency greater than the mean defection frequency $\left<f_{\textrm{C}}\right> > \left<f_{\textrm{D}}\right>$) under death-birth updating, again, from the perspective of effective game. However, the effects of game transitions can not be extended to the snowdrift game (see Fig.~S8). Specifically, the benefit in game 2, $b_2$, has no impact on evolutionary outcomes, as if the change of games does not occur (refer to Supplementary Note~6).

We have developed a mathematical framework and deduced a correspondingly general condition predicting evolutionary outcomes under varying environments. Several challenges and possible future research directions suggested by this work are: (i) In the main text, we assume that the game transitions are global, in which the games played by any two players have the potential to undergo an update. Compared to global transitions, the local game transition~\cite{su2019evolutionary} refers to the situation in which only those individuals competing for reproduction are more willing to change the environments they face. Therefore, it remains an open question how local game transitions affect the evolutionary outcomes for heterogeneous population structures. (ii) Though from the perspective of an ``effective game'', we have obtained a preliminary framework to calculate the fixation probability of cooperation for evolution in stochastic games, which successfully incorporates the heterogeneity of spatial structures. However, in general, such an approach is indeed an approximate method that captures part of the main features of stochastic games. A more accurate and realistic method could be the scope of subsequent research endeavors.

\section{Methods}
In this section, we briefly summarize our mathematical analysis for strategy evolution, which can be applied to a wide range of game transitions, and we refer to Supplementary Notes~1 and 2 for detailed derivations. 

We consider a weighted and undirected population structure $\mathcal{G}$ with edge weights $\mathcal{W} = \{w_{ij}\}_{i,j \in \mathcal{G}}$. All edges are supposed to be symmetric, i.e., $w_{ij} = w_{ji}$, and without self-loops. The weighted degree of individual $i$ is defined as $w_{i} = \sum_{j \in \mathcal{G}} w_{ij}$. Under death-birth updating, $\pi_i = w_{i} / \sum_{j \in \mathcal{G}} w_j$ represents the reproductive value of $i$, which measures the contribution of $i$ to the gene pool in the neutral evolution. 

Generally, the evolutionary dynamics can be modeled theoretically via random walks and coalescent theory on networks~\cite{allen2017evolutionary}. We follow the conventional definition of random walks on $\mathcal{G}$, i.e., let $p^{(1)}_{ij} = w_{ij} / w_{i}$ {($p^{(n)}_{ij}$)} denote the probability of starting from $i$ and terminating at $j$ after {one step ($n$ steps).} In the following, we show how to utilize the coalescent theory~\cite{allen2024coalescent}, including two and three-dimensional coalescing random walks, to calculate the desired critical benefit-to-cost ratio (details in Supplementary Notes~1 and 2).

\subsection{Fixation Probabilities for Arbitrary Pairwise Games}
For an arbitrary $2$-player game with the following payoff structure
\begin{equation}
    \begin{pNiceMatrix}[first-row,first-col]
        & \textrm{C} & \textrm{D} \\
    \textrm{C} & a_{11} & a_{12} \\
    \textrm{D} & a_{21} & a_{22} 
    \end{pNiceMatrix},
    \label{arbitrary matrix}
\end{equation}
the general expression of the fixation probability for cooperation is  
\begin{equation}
\frac{d}{d\delta}\Bigg\vert_{\delta =0} \rho_{\textrm{C}} = \frac{1}{N} \left[ \substack{ \left( a_{12}-a_{22} \right) \eta_{(2)} + \left( a_{21}-a_{22} \right)\left( \eta_{(3)} - \eta_{(1)} \right) \\+ \left( a_{11}-a_{12}-a_{21}+a_{22} \right) \left(\Lambda_{(2)} - \eta_{(1)} \right) } \right],     
\label{general fixation}
\end{equation}
where $\eta_{(n)} = \sum_{i,j \in \mathcal{G}} \pi_i p^{(n)}_{ij} \eta_{ij}$ and $\Lambda_{(n)} = \sum_{i,j,k \in \mathcal{G}} \pi_i p^{(n)}_{ij}p^{(1)}_{jk} \eta_{ijk}$. Furthermore, $\eta_{ij}$ and $\eta_{ijk}$, representing the expected coalescence time from subsets $\{i,j\}$ and $\{i,j,k\}$ belonging to $\mathcal{G}$, can be solved by the general system of linear equations ($I = (i_{1}, \dots, i_{m})$ with $|I| \geq 2$)
\begin{equation}
    \eta_{I} = \frac{1}{m} + \frac{1}{m} \sum_{y \in \mathcal{G}} \left( p^{(1)}_{i_{1}y}\eta_{y,\dots,i_{m}} + \dots + p^{(1)}_{i_{m}y}\eta_{i_{1},\dots,y} \right)
    \label{eta I}
\end{equation}
in polynomial time. Furthermore, Eq.~\ref{eta I} involves a system of linear equations of size $O(N^{|I|})$. Concretely, its size is $O(N^2)$ (resp. $O(N^3)$) for $|I|=2$, $\left\{\eta_{ij}\right\}_{i,j \in \mathcal{G}}$ (resp. for $|I|=3$, $\left\{\eta_{ijk}\right\}_{i,j,k \in \mathcal{G}}$).

Eq.~\ref{general fixation} generalizes the results for additive games, where the corresponding payoff function is of degree one~\cite{mcavoy2021fixation}, indicating $a_{11}+a_{22} = a_{12}+a_{21}$ (a property known as ``equal gains from switching''~\cite{allen2017evolutionary}). This property tells us that the Structure Coefficient Theorem~\cite{tarnita2009strategy} {$\sigma a_{11} + a_{12} > a_{21} + \sigma a_{22}$} implies $\rho_{\textrm{C}} > 1/N > \rho_{\textrm{D}}$. However, in a general case, it cannot provide the information of whether $\rho_{\textrm{C}} > 1/N$~\cite{taylor2007inclusive}. In particular, {DG} is one of the additive games, but SG is not. Therefore, by substituting Eq.~\ref{arbitrary matrix} with the payoff structure of DG, we can obtain the fixation probability for cooperation in~\cite{allen2017evolutionary}. 

\subsection{Critical Ratio for DG and SG with Game Transitions}
We proceed with the ``effective game'' approach, proposed and applied in~\cite{su2019evolutionary}, to approximate the evolutionary process with game transitions. It refers to the situation where the game transitions seemingly make all individuals accumulate an ``effective payoff'' during strategy evolution. We consider the endogenous transition for further analysis. In DG, cooperators receive the effective payoff $b_1-c$ for mutual cooperation; the unilateral defector obtains the temptation for defection $b_2$, while its opponent loses $c$ payoff for cooperation. Last, there is zero payoff for mutual defectors. Hence, the effective payoff structure of DG and SG (analysis is analogous) can be expressed as 
\begin{equation}
    \begin{pNiceMatrix}[first-row,first-col]
        & \textrm{C} & \textrm{D} \\
    \textrm{C} & b_1-c &   -c \\
    \textrm{D} & b_2 & 0 
    \end{pNiceMatrix},
    \label{effective DG}
\end{equation}
and
\begin{equation}
    \begin{pNiceMatrix}[first-row,first-col]
        & \textrm{C} & \textrm{D} \\
    \textrm{C} & b_1 - \frac{c}{2} &  b_2 - c \\
    \textrm{D} & b_2 & 0 
    \end{pNiceMatrix},
    \label{effective SG}
\end{equation}
respectively. It is worth noting that the static payoff structure might not be limited to the mentioned games. Therefore, substituting Eqs.~\ref{effective DG} and \ref{effective SG} into Eq.~\ref{general fixation}, the corresponding critical benefit-to-cost ratios for DG and SG are then given by
\begin{equation}
\left( \frac{b_1}{c}\right)^*_{\text{DG}} = \frac{\eta_{(2)}}{\eta_{(3)} - \eta_{(1)}} - \frac{\Lambda_{(2)}-\eta_{(3)}}{\eta_{(3)}-\eta_{(1)}} \cdot \frac{\Delta b}{c},
\label{DG ratio}
\end{equation} 
and
\begin{equation}
\begin{split}
\left( \frac{b_1}{c} \right)^{*}_{\text{SG}} &= \frac{ \Lambda_{(2)} - \eta_{(1)} - 2\eta_{(2)} }{2 \left(\Lambda_{(2)} - \eta_{(2)} - \eta_{(3)} \right)} \\
& \quad + \frac{ \eta_{(1)} + \eta_{(2)} + \eta_{(3)} - 2\Lambda_{(2)}}{\Lambda_{(2)} - \eta_{(2)} - \eta_{(3)}} \cdot  \frac{\Delta b}{c},     
\end{split}
\label{SG ratio}
\end{equation}
respectively, where $\Delta b = b_1 - b_2$. The above two equations give the critical ratio for evolution with game transitions in a general case under different social dilemmas. Furthermore, the first term on the right-hand side of the aforementioned equations is the critical benefit-to-cost ratio in every single game, and the second term encapsulates the influence of game transitions on the threshold for the evolution of cooperative behavior. In particular, $\Delta b$ quantifies the influence of variations across different games on the critical ratio. 

For the general transition rule, analytical analysis and detailed derivations can be found in Supplementary Notes~1 and 2.

\subsection{Relative measure of success}
{We have discussed the absolute measure of success, $\rho_{\textrm{C}}>1/N$, and analysis to the relative measure, $\rho_{\textrm{C}}>\rho_{\textrm{D}}$, is straightforward, by replacing $\eta_{I}$ (Eq.~\ref{eta I}) to $\tau_{I}$, where }
\begin{equation}
   \tau_{I} = 1 + \frac{1}{m} \sum_{y \in \mathcal{G}} \left( p^{(1)}_{i_{1}y}\tau_{y,\dots,i_{m}} + \dots + p^{(1)}_{i_{m}y}\tau_{i_{1},\dots,y} \right). 
\end{equation}
Following the definitions of $\eta_{(n)}$ and $\Lambda_{(n)}$, we denote $\tau_{(n)} = \sum_{i,j \in \mathcal{G}} \pi_i p^{(n)}_{ij} \tau_{ij}$ and $\Gamma_{(n)} = \sum_{i,j,k \in \mathcal{G}} \pi_i p^{(n)}_{ij}p^{(1)}_{jk} \tau_{ijk}$. We find that the following equality holds
\begin{equation}
    \Gamma_{(2)} = \frac{\tau_{(1)} + \tau_{(2)} + \tau_{(3)}}{2}\,.
    \label{equality}
\end{equation}
Replacing $\eta_{(n)}$ and $\Lambda_{(n)}$ with $\tau_{(n)}$ and $\Gamma_{(n)}$ in Eq.~\ref{general fixation} respectively and applying the above equality, we get
\begin{equation}
\begin{split}
    &(a_{11}+a_{12}-a_{21}-a_{22})\tau_{(2)} \\&+ (a_{11}-a_{12}+a_{21}-a_{22})\left( \tau_{(3)} - \tau_{(1)} \right) > 0\,,
\end{split}
\end{equation}
which agrees with the result obtained in~\cite{allen2019mathematical}.

Furthermore, Eq.~\ref{equality} can be extended to a general case, where $\Gamma_{(n)}$ can be expressed by only finite terms of $\tau_{(n)}$ (refer to Supplementary Note~8 for technical details). We leave the proof of the general case for future investigation. Consequently, the critical benefit-to-cost ratio for $\rho_{\textrm{C}}>\rho_{\textrm{D}}$ can be further simplified. Therefore, for the deterministic game transition, we have the following form for DG

\begin{equation}
    \left( \frac{b_1}{c}\right)^* = \frac{\tau_{(2)}}{\tau_{(3)} - \tau_{(1)}} - \frac{\tau_{(1)} + \tau_{(2)} - \tau_{(3)}}{2\left( \tau_{(3)} - \tau_{(1)} \right)} \cdot \frac{\Delta b}{c}\\,
    \label{relative DG}
\end{equation}
and

\begin{equation}
   \left( \frac{b_1}{c} \right)^{*} = \frac{ \tau_{(1)} + 3\tau_{(2)} - \tau_{(3)}}{2 \left( -\tau_{(1)} + \tau_{(2)} + \tau_{(3)} \right)}
    \label{relative SG}
\end{equation}
for SG. Note that, different from Eq.~\ref{relative DG} in the context of DG, Eq.~\ref{relative SG} does not include the term $\Delta b$, i.e., $b_2$ has no impact on the critical benefit-to-cost ratio in SG.

\section*{Data Availability}
All empirical network datasets utilized in this paper are freely and publicly available at \url{http://networkrepository.com} (see~\cite{rossi2015network}) and \url{https://networkx.org} (see~\cite{hagberg2008exploring}).

\section*{Code Availability}
All computational simulations, numerical calculations, and data analysis were performed in Python 3.7.4. All codes developed in this study have been deposited into a publicly available GitHub repository at \url{https://github.com/YujiZhangSWU/fixation-and-MSS}.

\section*{Acknowledgment}
This work was supported by the Chongqing Social Science Planning Project under Grant No. 2025NDQN41, by the Natural Science Foundation of Chongqing under Grant CSTB2025YITP-QCRCX0007, by the Slovenian Research and Innovation Agency under Grant No. P1-0403, by the National Research, Development and Innovation Office (NKFIH) under Grant No. K142948.

\section*{Author contributions}
Y.Z., M.F., Q.L., M.P., and A.S. conceived of the study. Y.Z. designed the methodology. Y.Z., M.F., Q.L., M.P., and A.S. conducted the investigation. Y.Z. implemented the visualization. Y.Z., M.F., Q.L., M.P., and A.S. wrote and edited the manuscript.

\section*{Competing interests}
The authors declare no competing interests.


\clearpage
\onecolumn
\begin{center}
\Large\textbf{Supplementary Information for \\ Effects of Stochastic Games on Evolutionary Dynamics in Structured Populations} 
\end{center}

\renewcommand{\thesection}{Supplementary Note \arabic{section}}

\section{Modeling Evolutionary Dynamics}
The undirected and symmetric networked population, with size $N$ and without self-loops, can be described by $\mathcal{G} = (\mathcal{N},\mathcal{E}, \mathcal{W})$, where $\mathcal{N}$, $\mathcal{E}$, and $\mathcal{W}$ are the corresponding node set, edge set, and weight set, respectively. In each time step, each individual has two optional strategies: cooperation ($\textrm{C}$) or defection ($\textrm{D}$), and accumulates its payoff from all game interactions. Specifically, we choose the averaged payoffs, one of the two main conventions for aggregating the payoffs, for further analysis. The state of the process can be tracked as a binary vector $\mathbf{s} = \{s_i\}_{i \in \mathcal{G}} = \{0,1\}^{N}$, where $s_i = 1$ (resp. $s_i = 0$) means individual $i$ takes cooperation (resp. defection). 

Concerning the strategy replacement rule, we take the death-Birth update rule \cite{ohtsuki2006simple} into consideration. Concretely, one individual is randomly chosen to be replaced in each time step, and its neighbors compete for the empty site in proportion to their fitness, where the fitness function is $F_i(\mathbf{s}) = 1 + \delta f_i(\mathbf{s}) + O(\delta^2)$ and $\delta$ represents the selection strength.

The two states $\mathbf{C}:= \{1,1,\dots,1\}$ and $\mathbf{D}:= \{0,0,\dots,0\}$ are the only two absorbing ones, excluding the existence of mutation, and all other states are transient during evolution. In this study, we focus on the fixation probability of cooperation (resp. defection) $\rho_{\textrm{C}}$ (resp. $\rho_{\textrm{D}}$) under weak selection ($0 < \delta \ll 1$) \cite{wu2010universality} and adopt the absolute measure ($\rho_{\textrm{C}} > 1/N$) and relative measure of success ($\rho_{\textrm{C}} > \rho_{\textrm{D}}$) as the criterion for the dominance of cooperation over neutral drift and defection.

In the following part, we will proceed with the workflow of McAvoy and Allen \cite{mcavoy2021fixation} for analyzing mutation-free evolutionary dynamics for arbitrary pairwise games. It is a straightforward derivation; therefore, only the main modifications are provided.

We consider a two-strategy matrix game with a general payoff structure
\begin{equation}
    \begin{pNiceMatrix}[first-row,first-col]
        & \textrm{C} & \textrm{D} \\
    \textrm{C} & a_{11} & a_{12} \\
    \textrm{D} & a_{21} & a_{22} 
    \end{pNiceMatrix}
\end{equation}
involving four generic parameters $a_{11}$, $a_{12}$, $a_{21}$, and $a_{22}$. Since adding one constant to each element in the payoff matrix does not affect the fixation probability, it is equivalent to
\begin{equation}
    \begin{pNiceMatrix}[first-row,first-col]
        & \textrm{C} & \textrm{D} \\
    \textrm{C} & a_{11}-a_{22} & a_{12}-a_{22} \\
    \textrm{D} & a_{21}-a_{22} & 0 
    \end{pNiceMatrix}.
\end{equation}
Based on it, the payoff function of $i$ can be expressed as 
\begin{equation}
f_i(\mathbf{s}) = C s_i + B\sum_{k \in \mathcal{G}} p^{(1)}_{ik}s_k + Ds_i\sum_{k \in \mathcal{G}} p^{(1)}_{ik} s_k,
\end{equation}
where $B=a_{21}-a_{22}$, $C=a_{12}-a_{22}$, and $D=a_{11}-a_{12}-a_{21}+a_{22}$. The general payoff function indicates that such a game is not additive \cite{mcavoy2021fixation}, since $f_i(\mathbf{s})$ is of degree two (nonlinear) in $\mathbf{s}$ for every $i \in \mathcal{G}$, unless $D = 0$.

Differentiating the marginal probability $e_{ij}(\mathbf{s})$ that $i$ transmits its strategy to $j$ with respect to $\delta$ at $\delta=0$ gives
\begin{equation}
\frac{d}{d\delta}\bigg|_{\delta=0} e_{ij}(\mathbf{s}) = \frac{p^{(1)}_{ji}}{N}\left(f_i(\mathbf{s}) - \sum_{l \in \mathcal{G}} p^{(1)}_{jl} f_l(\mathbf{s})\right). 
\end{equation}
Substituting 
\begin{alignat*}{1}
f_i(\mathbf{s}) - \sum_{l \in \mathcal{G}} p^{(1)}_{jl} f_l(\mathbf{s})  &= C s_i + B\sum_{k \in \mathcal{G}} p^{(1)}_{ik}s_k + Ds_i\sum_{k \in \mathcal{G}} p^{(1)}_{ik} s_k - \sum_{l \in \mathcal{G}} p^{(1)}_{jl} \left(C s_l + B\sum_{k \in \mathcal{G}} p^{(1)}_{lk}s_k + Ds_l\sum_{k \in \mathcal{G}} p^{(1)}_{lk} s_k \right)\\
&= C \sum_{k \in \mathcal{G}}\left(p^{(0)}_{ik} - p^{(1)}_{jk}\right)s_k + B \sum_{k \in \mathcal{G}} \left(  p^{(1)}_{ik} - p^{(2)}_{jk} \right)s_k + D \sum_{k,l \in \mathcal{G}} \left( p^{(0)}_{ik} p^{(1)}_{il} - p^{(1)}_{jk} p^{(1)}_{kl} \right) s_k s_l
\end{alignat*}
into the above equation yields 
\begin{equation}
\frac{d}{d\delta}\bigg|_{\delta=0}  e_{ij}(\mathbf{s}) = \sum_{k \in \mathcal{G}} c^{ij}_{k} s_{k} + \sum_{k,l \in \mathcal{G}} c^{ij}_{kl} s_{k}s_{l}\,,
\end{equation}
where  
\begin{equation}
    c_{k}^{ij} = \frac{p^{(1)}_{ji}}{N} \left(C \left(p^{(0)}_{ik} - p^{(1)}_{jk}\right) + B \left(  p^{(1)}_{ik} - p^{(2)}_{jk} \right) \right)
\end{equation}
and
\begin{equation}
    c_{kl}^{ij} =  \frac{p^{(1)}_{ji}}{N} \left( D \left( p^{(0)}_{ik} p^{(1)}_{il} - p^{(1)}_{jk} p^{(1)}_{kl} + p^{(0)}_{il} p^{(1)}_{ik} - p^{(1)}_{jl} p^{(1)}_{lk} \right) \right). 
\end{equation}
Furthermore, $(k)$ and $(k,l)$, representing the linear and quadratic coefficients, belong to 
\begin{equation}
    \mathcal{L}^{(1)} = \{(1),(2),\cdots,(N)\},
\end{equation}
and 
\begin{equation}
    \mathcal{L}^{(2)} = \{(1,2),\cdots,(1,N),(2,3),\cdots,(N-1,N)\},
\end{equation}
respectively.

We are now in a position to calculate $\rho_{\textrm{C}}$ with the framework proposed by McAvoy and Allen \cite{mcavoy2021fixation}. For any initial configuration $\bm{\xi} = \{\xi_1,\dots,\xi_N\}$, applying Theorem 2 of Ref. \cite{mcavoy2021fixation}, $\rho_{\textrm{C}}(\bm{\xi})$ is then given by the general expression
\begin{equation}
\begin{split}
\rho_{\textrm{C}}(\bm{\xi}) &= \widehat{\xi} + \delta \left( \sum_{i,j \in \mathcal{G}}\sum_{(k) \in \mathcal{L}^{(1)}} \pi_i c^{ji}_{k} \left(\eta^{\bm{\xi}}_{ik} - \eta^{\bm{\xi}}_{jk} \right) + \sum_{i,j \in \mathcal{G}}\sum_{(k,l) \in \mathcal{L}^{(2)}} \pi_i c^{ji}_{kl} \left(\eta^{\bm{\xi}}_{ikl} - \eta^{\bm{\xi}}_{jkl} \right) \right) + O(\delta^2) \\
&= \widehat{\xi} + \frac{\delta}{N} \left(\sum_{i,j,k \in \mathcal{G}} \pi_i p^{(1)}_{ij} \left( C \left(p^{(0)}_{jk} - p^{(1)}_{ik}\right) + B \left(  p^{(1)}_{jk} - p^{(2)}_{ik} \right) \right) \left(\eta^{\bm{\xi}}_{ik} - \eta^{\bm{\xi}}_{jk} \right) \right) \\
& \quad + \frac{\delta}{N} \left(D \sum_{i,j,k,l \in \mathcal{G}} \pi_i p^{(1)}_{ij} \left( p^{(0)}_{jk} p^{(1)}_{jl} - p^{(1)}_{ik} p^{(1)}_{kl} \right) \left(\eta^{\bm{\xi}}_{ikl} - \eta^{\bm{\xi}}_{jkl} \right) \right) + O(\delta^2) \\
&= \widehat{\xi} + \frac{\delta}{N} \left[ C \eta^{\bm{\xi}}_{(2)} + B\left( \eta^{\bm{\xi}}_{(3)} - \eta^{\bm{\xi}}_{(1)} \right) + D \left(- \sum_{i,j,k \in \mathcal{G}} \pi_i p^{(1)}_{ij}p^{(1)}_{jk} \eta_{jk} + \sum_{i,j,k \in \mathcal{G}} \pi_i p^{(2)}_{ij}p^{(1)}_{jk} \eta_{ijk} \right)\right] + O(\delta^2) \\
&= \widehat{\xi} + \frac{\delta}{N} \left[ C \eta^{\bm{\xi}}_{(2)} + B\left( \eta^{\bm{\xi}}_{(3)} - \eta^{\bm{\xi}}_{(1)} \right) + D \left(\Lambda^{\bm{\xi}}_{(2)} - \eta^{\bm{\xi}}_{(1)} \right)\right] + O(\delta^2)    
\label{fixation probability}
\end{split}
\end{equation}
where $\widehat{\xi} := \sum_{i \in \mathcal{G}} \pi_i \xi_i$, $\eta^{\bm{\xi}}_{(n)} := \sum_{i,j \in \mathcal{G}} \pi_i p^{(n)}_{ij} \eta^{\bm{\xi}}_{ij}$, and $\Lambda^{\bm{\xi}}_{(n)} := \sum_{i,j,k \in \mathcal{G}} \pi_i p^{(n)}_{ij}p^{(1)}_{jk} \eta^{\bm{\xi}}_{ijk}$. Simplification can be done with the application of the reversibility property $\pi_i p^{(n)}_{ij} = \pi_j p^{(n)}_{ji}$. Recall that $\pi_i$ means the reproductive value of $i$ and is uniquely determined by the linear equations given below \cite{allen2014measures,allen2019mathematical}.
\begin{equation}
\begin{split}
    \sum_{j=1}^{N}e^{\circ}_{ij} \pi_j &= \sum_{j=1}^{N}e^{\circ}_{ji} \pi_i \\
    \sum_{i=1}^{N} \pi_i &= 1
\end{split}
\end{equation}
where the superscript $\circ$ refers to the value is taken under neutral drift ($\delta = 0$).

In a degree-two process, only terms with the form $c^{ij}_{\emptyset}$, $c^{ij}_{k}s_k$ and $c^{ij}_{kl}s_ks_l$ exist. Therefore, under the uniform initialization, we have $\eta_{i}=0$ for $|I|=1$. 

For $|I|=2$, $\eta_{ij}$ is the unique solution to the equations for $i \neq j$,
\begin{equation}
\eta_{ij} = \frac{1}{2} + \frac{1}{2}\sum_{k\in \mathcal{G}} \left( p^{(1)}_{ik}\eta_{kj} + p^{(1)}_{jk} \eta_{ik} \right).
\end{equation}

For $|I|=3$, $\eta_{ijk}$ is the unique solution to the equations for $i \neq j \neq k \neq i$,
\begin{equation}
\eta_{ijk} = \frac{1}{3} + \frac{1}{3}\sum_{l\in \mathcal{G}} \left( p^{(1)}_{il} \eta_{ljk} + p^{(1)}_{jl} \eta_{ilk} + p^{(1)}_{kl} \eta_{ijl} \right).
\end{equation}

\section{Strategy Evolution in Stochastic Games}
As a conventional setting, we focus on a particular game called the donation game, of which the payoff structure is 
\begin{equation}
    \begin{pNiceMatrix}[first-row,first-col]
        & \textrm{C} & \textrm{D} \\
    \textrm{C} &  b-c &   -c \\
    \textrm{D} &  b & 0 
    \end{pNiceMatrix}.
\end{equation}
In the following part, we will apply the framework proposed in the previous section for analyzing mutation-free evolutionary dynamics with game transitions, by introducing the ``effective game'' method.

First, we denote $E = \{1,\dots,L\}$ as the state space of all possible $L$ game states during evolution. Hence, game transitions among $L$ states can be described by $\mathbf{P}^{[s]}$, that is
\begin{equation}
\mathbf{P}^{[0]} = 
\begin{pmatrix}
    p^{[0]}_{11}  & \cdots & p^{[0]}_{1L} \\
    \vdots  & \ddots  & \vdots \\
    p^{[0]}_{L1}  & \cdots & p^{[0]}_{LL} \\
\end{pmatrix},\, 
\mathbf{P}^{[1]} = 
\begin{pmatrix}
    p^{[1]}_{11}  & \cdots & p^{[1]}_{1L} \\
    \vdots  & \ddots  & \vdots \\
    p^{[1]}_{L1}  & \cdots & p^{[1]}_{LL} \\
\end{pmatrix},\,
\mathbf{P}^{[2]} = 
\begin{pmatrix}
    p^{[2]}_{11}  & \cdots & p^{[2]}_{1L} \\
    \vdots  & \ddots  & \vdots \\
    p^{[2]}_{L1}  & \cdots & p^{[2]}_{LL} \\
\end{pmatrix}
\end{equation}
where $s \in \{0,1,2\}$ is the number of cooperators in each dyadic interaction and $p^{[s]}_{ij}$ denotes the conditional probability that given a scenario in which there are $s$  $\textrm{C}$-players in total, players will switch to playing game $j$ in the subsequent time based on their current engagement with game $i$. To avoid confusion, we utilize the superscripts $[s]$ and $(n)$ to represent symbols related to $\mathbf{P}^{[s]}$ (e.g., the stationary distribution $\mathbf{u}^{[s]}$ of $\mathbf{P}^{[s]}$) and notations related to $n$-step random walk on $\mathcal{G}$ (e.g., the probability $p^{(n)}_{ij}$ that an $n$-step random walk from $i$ terminates at $j$), respectively.

Let $\mathbf{u}^{[s]} = \left( u^{[s]}_1,\dots,u^{[s]}_L \right)$, satisfying $\mathbf{u}^{[s]} = \mathbf{u}^{[s]} \mathbf{P}^{[s]}$, represent the stationary distribution of the Markov chain $\mathbf{P}^{[s]}$, which has only one recurrence class. We provide one approach to analyze a wide range of game transition patterns from the perspective of ``effective game'' \cite{su2019evolutionary}. It refers to the situation in which the game transitions seemingly make all individuals accumulate an ``effective payoff'' during strategy evolution, of which the payoff structure can be described as 
\begin{equation}
    \begin{pNiceMatrix}[first-row,first-col]
        & \textrm{C} & \textrm{D} \\
    \textrm{C} & \sum_{i=1}^{L} u_{i}^{[2]} b_i-c &   -c \\
    \textrm{D} & \sum_{i=1}^{L} u_{i}^{[1]} b_i & 0 
    \end{pNiceMatrix},
\end{equation}
indicating $B = \sum_{i=1}^{L} u_{i}^{[1]} b_i$, $C=-c$ and $D = \sum_{i=1}^{L} \left( u_{i}^{[2]} - u_{i}^{[1]} \right) b_i$. 

Hence, substituting $B$, $C$ and $D$ into Eq.~S\ref{fixation probability}, the critical ratio for $\rho_{\textrm{C}} > 1/N$ is then given by
\begin{equation}
\left( \frac{b_1}{c}\right)^* = \frac{\eta_{(2)}}{\eta_{(3)} - \eta_{(1)}} + 
\sum^{L}_{i=2}\left( \frac{\Lambda_{(2)}-\eta_{(1)}}{\eta_{(3)}-\eta_{(1)}}u^{[2]}_{i} - \frac{\Lambda_{(2)}-\eta_{(3)}}{\eta_{(3)}-\eta_{(1)}}u^{[1]}_{i}\right) \frac{\Delta b_{1i}}{c}
\label{general result}
\end{equation}
where $\Delta b_{1i} := b_1 - b_i$.

Moreover, we can derive the condition of $\rho_{\textrm{C}} > \rho_{\textrm{D}}$, given by
\begin{equation}
\left( \frac{b_1}{c}\right)^* = \frac{\tau_{(2)}}{\tau_{(3)} - \tau_{(1)}} + \sum^{L}_{i=2}\left( \frac{\Gamma_{(2)}-\tau_{(1)}}{\tau_{(3)}-\tau_{(1)}}u^{[2]}_{i} - \frac{\Gamma_{(2)}-\tau_{(3)}}{\tau_{(3)}-\tau_{(1)}}u^{[1]}_{i}\right) \frac{\Delta b_{1i}}{c}
\end{equation}
where $\tau_{(n)} := \sum_{i,j \in \mathcal{G}} \pi_i p^{(n)}_{ij} \tau_{ij}$ and $\Gamma_{(n)} := \sum_{i,j,k \in \mathcal{G}} \pi_i p^{(n)}_{ij}p^{(1)}_{jk} \tau_{ijk}$. For $|I|=2$, $\tau_{ij}$ is the unique solution to the equations for $i \neq j$,
\begin{equation}
\tau_{ij} = 1 + \frac{1}{2}\sum_{k\in \mathcal{G}} \left( p^{(1)}_{ik}\tau_{kj} + p^{(1)}_{jk} \tau_{ik} \right).
\end{equation}
For $|I|=3$, $\tau_{ijk}$ is the unique solution to the equations for $i \neq j \neq k$,
\begin{equation}
\tau_{ijk} = 1 + \frac{1}{3}\sum_{l\in \mathcal{G}} \left( p^{(1)}_{il} \tau_{ljk} + p^{(1)}_{jl} \tau_{ilk} + p^{(1)}_{kl} \tau_{ijl} \right).
\end{equation}
Specifically, when $\Delta b_{1i} = 0$ for each $i$, indicating such $L$ games are the same, we obtain the well-known result of Ref. \cite{allen2017evolutionary}, which applies to any connected graphs.

\section{Applications to Specific Examples}
Now we turn to some specific population structures and game transition modes.
\subsection{Complete Graphs}
For a complete network of size $N$ ($N \geq 3$), we have \cite{sheng2024strategy}
\begin{equation}
\begin{split}
    \eta_{ij} &= \frac{N-1}{2}, \\
     \eta_{ijk} &= \frac{2(N-1)}{3},    
\end{split}
\end{equation}
and (after straightforward calculations according to the definition of $\eta_{(n)}$ and $\Lambda_{(n)}$).
\begin{equation}
\begin{split}
    \eta_{(1)} &= \frac{N-1}{2}, \\
    \eta_{(2)} &= \frac{N-2}{2}, \\
    \eta_{(3)} &= \frac{N^2-3N+3}{2(N-1)}, \\
    \Lambda_{(2)} &= \frac{4N^2-10N+7}{6(N-1)}.    
\end{split}
\end{equation}

Then, the fixation probability of cooperation is given by 
\begin{equation}
\rho_{\textrm{C}} = \frac{1}{N} + \frac{\delta}{N} \left[ -C \left(\frac{N-2}{2}\right) - B\left(\frac{N-2}{2(N-1)} \right) + D \left(\frac{N^2-4N+4}{6(N-1)}\right) \right].
\end{equation}
Therefore, in sense of $\rho_{\textrm{C}} > 1/N$, the critical benefit-cost ratio is given by
\begin{equation}
\left( \frac{b_1}{c}\right)^* = -N + 1 + \sum_{i=2}^{L} \left( -\frac{N-2}{3} u^{[2]}_{i} + \frac{N+1}{3}u^{[1]}_{i} \right) \frac{\Delta b_{1i}}{c}\,.
\end{equation}

\subsection{Ceiling Fan Graphs}
Let us now consider the ``ceiling fan'' graph, a more complicated network topology, where an edge joins each of the $2n$ ($n \geq$ 2) leaves of a star to one other. Due to the symmetry, we can label the root (focal) node as $r$ and the other leaf nodes as $l$. Furthermore, we utilize the superscripts $+$ and $-$ to distinguish two leaves from the same triplet and subscripts $1$ and $2$ to represent two leaves from different triplets (e.g., $\eta_{l^{+}_{1}l^{-}_{1}}$, $\eta_{l_{1}l_{2}}$ and $\eta_{rl_{1}}$). 

According to Ref. \cite{sheng2024strategy}, we have 
\begin{equation}
\begin{split}
    \eta_{r l_{1}} &= \frac{8n-3}{2n+3}, \\
    \eta_{l^{+}_{1} l^{-}_{1}} &= \frac{5n}{2n+3}, \\ 
    \eta_{l_{1} l_{2}} &= \frac{10n}{2n+3}, \\
    \eta_{r l^{+}_{1} l^{-}_{1}} &= \frac{2\left( 205n^3 + 270n^2 - 79n - 6\right)}{3\left(2n+3\right)\left(14n^2+23n+2\right)}, \\
    \eta_{r l_{1} l_{2}} &= \frac{4\left( 119n^3 + 174n^2 - 20n - 3\right)}{3\left(2n+3\right)\left(14n^2+23n+2\right)}, \\
    \eta_{l^{+}_{1} l^{-}_{1} l_{2}} &= \frac{2\left( 79n^3 + 128n^2 + 8n\right)}{\left(2n+3\right)\left(14n^2+23n+2\right)},\\ 
    \eta_{l_{1} l_{2} l_{3}} &= \frac{2\left( 266n^3 + 436n^2 + 33n\right)}{3\left(2n+3\right)\left(14n^2+23n+2\right)},
\end{split}
\end{equation}
and
\begin{equation}
\begin{split}
    \eta_{(1)} &= \frac{21n-6}{3\left(2n+3\right)}, \\ 
    \eta_{(2)} &= \frac{36n-21}{6\left(2n+3\right)}, \\
    \eta_{(3)} &= \frac{93n-48}{12\left(2n+3\right)}, \\
    \Lambda_{(2)} &= \frac{4816n^4+5969n^3-2686n^2-482n-12}{36n\left(2n+3\right)\left(14n^2+23n+2\right)}.
\end{split}
\end{equation}

Therefore, the critical benefit-cost ratio for $\rho_{\textrm{C}} > 1/N$ is
\begin{equation}
\left( \frac{b_1}{c}\right)^* = \frac{2\left(12n-7\right)}{3n-8} + \sum_{i=2}^{L} \left(\frac{\left(\substack{1288n^4+1181n^3\\-1534n^2-338n\\-12}\right)}{\left(\substack{378n^4-387n^3\\-1602n^2-144n}\right)}u^{[2]}_{i} 
- \frac{\left(\substack{910n^4+1568n^3\\+68n^2-194n\\-12}\right)}{\left(\substack{378n^4-387n^3\\-1602n^2-144n}\right)}u^{[1]}_{i}\right) \frac{\Delta b_{1i}}{c}.
\end{equation}

Letting $n \rightarrow \infty$, we have
\begin{equation}
\left( \frac{b_1}{c}\right)^* = 8 + \sum_{i=2}^{L} \left(\frac{92}{27}u^{[2]}_{i} 
- \frac{65}{27}u^{[1]}_{i}\right) \frac{\Delta b_{1i}}{c}.
\end{equation}

\subsection{Conjoining Two Star Graphs}
This subsection considers a network by conjoining two identical star graphs of size $n$ ($n \geq 4$). Two root (focal) nodes are selected in each sub-network to be the hub, and two hubs are connected with an edge. The remaining nodes are called leaves. For simplicity, we utilize the superscripts $+$ and $-$ to distinguish two leaves from the same star graph and the subscripts $1$ and $2$ to represent nodes from different subgraphs. 

For $|I|=2$, the system of linear equations can be expressed as
\begin{equation}
\begin{split}
    \eta_{r_1r_2} &= \frac{1}{2} + \frac{1}{2} \left( \frac{2\left(n-1\right)}{n}\eta_{r_1l_2} \right), \\
    \eta_{r_1l_1} &= \frac{1}{2} + \frac{1}{2} \left( \frac{n-2}{n}\eta_{l^+_1l^-_1} + \frac{1}{n}\eta_{r_1l_2} \right), \\
    \eta_{r_1l_2} &= \frac{1}{2} + \frac{1}{2} \left( \frac{1}{n}\eta_{r_1l_1} + \frac{n-1}{n}\eta_{l_1l_2}  + \eta_{r_1r_2} \right), \\
    \eta_{l^+_1l^-_1} &= \frac{1}{2} + \eta_{r_1l_1}, \\
    \eta_{l_1l_2} &= \frac{1}{2} + \eta_{r_1l_2}.    
\end{split}
\end{equation}
Solving the above system of linear equations gives
\begin{equation}
\begin{split}
    \eta_{r_1r_2} &= \frac{4n^3+8n^2-11n+4}{2n\left(2n+3\right)}, \\
    \eta_{r_1l_1} &= \frac{10n-5}{2\left(2n+3\right)}, \\
    \eta_{r_1l_2} &= \frac{2n^2+5n-2}{2n+3}, \\
    \eta_{l^+_1l^-_1} &= \frac{6n-1}{2n+3}, \\
    \eta_{l_1l_2} &= \frac{4n^2+12n-1}{2\left(2n+3\right)}.
\end{split}
\end{equation}

For $|I|=3$, the system of linear equations becomes
\begin{equation}
\begin{split}
    \eta_{r_1l^+_1l^-_1} &= \frac{1}{3} + \frac{1}{3} \left( \frac{1}{n}\eta_{r_1l^+_2l^-_2} + \frac{2}{n}\eta_{l^+_1l^-_1} 
                        + \frac{n-3}{n} \eta_{l^+_1l^-_1l^*_1} + 2 \eta_{r_1l_1}  \right), \\
    \eta_{r_1r_2l_1} &= \frac{1}{3} + \frac{1}{3} \left( \frac{2}{n}\eta_{r_1l_2} + \frac{n-2}{n}\eta_{r_1l^+_2l^-_2} 
                        + \frac{1}{n}\eta_{r_1l_1} + \frac{n-1}{n} \eta_{r_1l_1l_2} + \eta_{r_1r_2}\right), \\
    \eta_{r_1l_1l_2} &= \frac{1}{3} + \frac{1}{3} \left( \frac{1}{n}\eta_{r_1l_1l_2} + \frac{1}{n}\eta_{l_1l_2} + \frac{n-2}{n}\eta_{l^+_1l^-_1l_2} 
                        + \eta_{r_1l_2} + \eta_{r_1r_2l_1} \right), \\
    \eta_{r_1l^+_2l^-_2} &=  \frac{1}{3} + \frac{1}{3} \left( \frac{1}{n}\eta_{r_1l^+_1l^-_1} + \frac{n-1}{n}\eta_{l^+_1l^-_1l_2} + 2\eta_{r_1r_2l_1} \right), \\
    \eta_{l^+_1l^-_1l_2} &= \frac{1}{3} + \frac{1}{3} \left( 2\eta_{r_1l_1l_2} + \eta_{r_1l^+_2l^-_2} \right), \\
    \eta_{l^+_1l^-_1l^*_1} &= \frac{1}{3} + \eta_{r_1l^+_1l^-_1}.
\end{split}
\end{equation}
Solving the system of linear equations gives
\begin{equation}
\begin{split}
    \eta_{r_1l^+_1l^-_1} &= \frac{616 n^{4} + 2082 n^{3} + 927 n^{2} - 846 n + 81}{12 \left(2 n + 3\right) \left(7 n^{3} + 27 n^{2} + 21 n - 3\right)}, \\
    \eta_{r_1r_2l_1} &= \frac{168 n^{6} + 1110 n^{5} + 2081 n^{4} + 227 n^{3} - 999 n^{2} + 291 n - 18}{12 n \left(2 n + 3\right) \left(7 n^{3} + 27 n^{2} + 21 n - 3\right)}, \\
    \eta_{r_1l_1l_2} &= \frac{56 n^{5} + 398 n^{4} + 845 n^{3} + 341 n^{2} - 240 n + 30}{4 \left(2 n + 3\right) \left(7 n^{3} + 27 n^{2} + 21 n - 3\right)}, \\
    \eta_{r_1l^+_2l^-_2} &= \frac{56 n^{5} + 384 n^{4} + 804 n^{3} + 299 n^{2} - 264 n + 21}{4 \left(2 n + 3\right) \left(7 n^{3} + 27 n^{2} + 21 n - 3\right)}, \\
    \eta_{l^+_1l^-_1l_2} &= \frac{168 n^{5} + 1236 n^{4} + 2794 n^{3} + 1473 n^{2} - 516 n + 45}{12 \left(2 n + 3\right) \left(7 n^{3} + 27 n^{2} + 21 n - 3\right)}, \\
    \eta_{l^+_1l^-_1l^*_1} &= \frac{224 n^{4} + 794 n^{3} + 473 n^{2} - 206 n + 15}{4 \left(2 n + 3\right) \left(7 n^{3} + 27 n^{2} + 21 n - 3\right)}.
\end{split}
\end{equation}

Consequently, we have
\begin{equation}
\begin{split}
    \eta_{(1)} &= \frac{24 n^{3} - 22 n^{2} - n + 4}{2 n \left(4 n^{2} + 4 n - 3\right)}, \\
    \eta_{(2)} &= \frac{10 n^{3} - 13 n^{2} + n + 2}{n \left(4 n^{2} + 4 n - 3\right)}, \\
    \eta_{(3)} &= \frac{32 n^{5} - 38 n^{4} + 5 n^{3} + 12 n^{2} - 10 n + 4}{2 n^{3} \left(4 n^{2} + 4 n - 3\right)}, \\
    \Lambda_{(2)} &= \frac{1540 n^{8} + 4008 n^{7} - 1877 n^{6} - 4501 n^{5} + 2236 n^{4} + 238 n^{3} - 513 n^{2} + 483 n - 54}{12 n^{3} \left(2 n - 1\right) \left(2 n + 3\right) \left(7 n^{3} + 27 n^{2} + 21 n - 3\right)}.   
\end{split}
\end{equation}

Then, the critical benefit-cost ratio for $\rho_{\textrm{C}} > 1/N$ is given by
\begin{equation}
\left( \frac{b_1}{c}\right)^* = \frac{n^{2} \left(10 n^{2} - 3 n - 2\right)}{\left(n - 1\right) \left(4 n^{3} - n + 2\right)} + \sum_{i=2}^{L} \left( \frac{ \left( \substack{ 532 n^{7} + 1576 n^{6} + 281 n^{5}\\ - 1022 n^{4} + 296 n^{3} + 12 n^{2}\\ - 429 n + 54 } \right)}{12 \left( \substack{ 28 n^{7} + 80 n^{6} - 31 n^{5}\\ - 102 n^{4} + 58 n^{3} + 12 n^{2}\\ - 51 n + 6 } \right)} u^{[2]}_{i} - \frac{\left(\substack{196 n^{7} + 616 n^{6} + 653 n^{5}\\ + 202 n^{4} - 400 n^{3} - 132 n^{2}\\ + 183 n - 18} \right)}{12 \left( \substack{ 28 n^{7} + 80 n^{6} - 31 n^{5}\\ - 102 n^{4} + 58 n^{3} + 12 n^{2}\\ - 51 n + 6 } \right)} u^{[1]}_{i} \right) \frac{\Delta b_{1i}}{c}.   
\end{equation}
Letting $n \to \infty$, we have 
\begin{equation}
\left( \frac{b_1}{c}\right)^* = \frac{5}{2} + \sum_{i=2}^{L} \left(\frac{19}{12}u^{[2]}_{i} - \frac{7}{12}u^{[1]}_{i}\right) \frac{\Delta b_{1i}}{c}.
\end{equation}

\subsection{Exogenous Game Transitions}
The exogenous game transition refers to the transition that is independent of the strategy evolution, indicating that $\mathbf{P}^{[0]} = \mathbf{P}^{[1]} = \mathbf{P}^{[2]}$. Without loss of generality, let $\mathbf{u}^{[s]} = (u_1,\dots,u_L)$ denote the stationary distribution of $\mathbf{P}^{[s]}$ for $s \in \{0,1,2\}$, where $\sum_{i=1}^{L} u_i=1$. Therefore, the critical ratio for $\rho_{\textrm{C}} > 1/N$ is given by
\begin{equation}
    \left(\frac{b_1}{c} \right)^{*} = \frac{\eta_{(2)}}{\eta_{(3)} - \eta_{(1)}} + \sum_{i=2}^{L} u_i \frac{\Delta b_{1i}}{c},
\end{equation}
which simultaneously implies the condition of the relative measure of success for cooperation.

\subsection{Endogenous Game Transitions}
The endogenous game transition indicates that the transition depends on the strategy evolution. For that purpose, we focus on one classic game transition pattern, i.e., the deterministic state-independent game transition, which can be described as  
\begin{equation}
    \begin{matrix}
    \mathbf{P}^{[0]}=\begin{bmatrix}
    0  & 1 \\ 
    0  & 1 
    \end{bmatrix}, \,
    \mathbf{P}^{[1]} =   
    \begin{bmatrix}
    0 & 1 \\ 
    0 & 1
    \end{bmatrix}, \,
    \mathbf{P}^{[2]} = 
    \begin{bmatrix}
    1 & 0 \\ 
    1 & 0
    \end{bmatrix}
    \end{matrix}.
\end{equation}
Therefore, the critical ratio for $\rho_{\textrm{C}} > 1/N$ is 
\begin{equation}
\left( \frac{b_1}{c}\right)^* = \frac{\eta_{(2)}}{\eta_{(3)} - \eta_{(1)}} - \left( \frac{\Lambda_{(2)}-\eta_{(3)}}{\eta_{(3)}-\eta_{(1)}} \right) \frac{\Delta b_{12}}{c},
\end{equation}
and the critical ratio for $\rho_{\textrm{C}} > \rho_{\textrm{D}}$ becomes
\begin{equation}
\left( \frac{b_1}{c}\right)^* = \frac{\tau_{(2)}}{\tau_{(3)} - \tau_{(1)}} - \left( \frac{\Gamma_{(2)}-\tau_{(3)}}{\tau_{(3)}-\tau_{(1)}} \right) \frac{\Delta b_{12}}{c}.
\label{end critical ratio}
\end{equation}

\section{Variants of Model}
Indeed, our framework can be applied to other games, not confined to the donation game. In this section, we take the public goods game with second-order interactions and the snowdrift game as two examples, though the public goods game is more suitable for modeling higher-order interactions, as analyzed in Ref. \cite{sheng2024strategy}.
\subsection{Transitions among Public Goods Game}
First, we consider the public goods game without game transitions. For the public goods game with the payoff matrix \cite{sheng2024strategy}
\begin{equation}
    \begin{pNiceMatrix}[first-row,first-col]
        & \textrm{C} & \textrm{D} \\
    \textrm{C} & -c+\frac{(\delta_{2}+1)b}{2} & -c+\frac{b}{2} \\
    \textrm{D} & \frac{b}{2} & 0 
    \end{pNiceMatrix}
\end{equation}
where $0 < \delta_{2} < 1$ (resp. $\delta_{2} \geq 1$) is the discounting (resp. synergy) factor. 

According to the payoff structure, the payoff function can be expressed as 
\begin{alignat*}{1}
f_i(\mathbf{s}) &= \left( -c+\frac{(\delta_{2}+1)b}{2} \right) s_i s_i^{(1)} + \left( -c+\frac{b}{2} \right) s_i (1-s_i^{(1)}) + \frac{b}{2} (1-s_i)s_i^{(1)} \\
&= -C s_i + B\sum_{k \in \mathcal{G}} p^{(1)}_{ik}s_k + Ds_i\sum_{k \in \mathcal{G}} p^{(1)}_{ik} s_k
\end{alignat*}
where $B = \frac{b}{2}$, $C = c-\frac{b}{2}$ and $D = \frac{(\delta_{2}-1)b}{2}$.

Consequently, the critical ratio for $\rho_{\textrm{C}} > 1/N$ is given by 
\begin{equation}
\left( \frac{b}{c} \right)^* = \frac{2 \eta_{(2)}}{\eta_{(2)} + \eta_{(3)} - \eta_{(1)} + \left( \delta_{2} - 1 \right)\left( \Lambda_{(2)} - \eta_{(1)}\right)}.
\end{equation}
Therefore, on a complete network of size $N \geq 3$, the critical ratio is
\begin{equation}
\left( \frac{b}{c}\right)^* = \frac{6\left( N - 1 \right)}{\left( \delta_{2} + 2\right) \left( N - 2\right)},
\end{equation}
which meets the result in Ref. \cite{sheng2024strategy} for accumulated payoffs, since averaged payoffs are equivalent to accumulated payoffs on complete networks.

Now we turn to the case in which transitions are among public good games, which can lead to the change of $b$, denoted by $b_i$ for $i \in \{ 1,2,\dots,L \}$. Hence, the ``effective'' payoff structure can be described as 
\begin{equation}
    \begin{pNiceMatrix}[first-row,first-col]
        & \textrm{C} & \textrm{D} \\
    \textrm{C} & -c + \sum_{i=1}^{L} u_{i}^{[2]} \frac{(\delta_{2} + 1)b_i}{2} & -c + \sum_{i=1}^{L} u_{i}^{[1]} \frac{b_i}{2} \\
    \textrm{D} & \sum_{i=1}^{L} u_{i}^{[1]} \frac{b_i}{2}  & 0 
    \end{pNiceMatrix}.
\end{equation}

According to the payoff structure, the coefficients of its payoff function are
\begin{equation}
\begin{split}
    B &= \sum_{i=1}^{L} u_{i}^{[1]} \frac{b_i}{2}, \\
    C &= \sum_{i=1}^{L} u_{i}^{[1]} \frac{b_i}{2} - c, \\
    D &= \sum_{i=1}^{L} u_{i}^{[2]} \frac{(\delta_{2} + 1)b_i}{2} - \sum_{i=1}^{L} u_{i}^{[1]} b_i.
\end{split}
\end{equation}

As a consequence, the critical ratio for $\rho_{\textrm{C}} > 1/N$ is given by 
\begin{equation}
\left( \frac{b_1}{c} \right)^* = \frac{2\eta_{(2)}}{\Xi} + \sum_{i=2}^{L}\left( \left( \delta_{2}+1 \right) \frac{ \Lambda_{(2)} - \eta_{(1)} }{\Xi} u_{i}^{[2]} - \frac{ 2\Lambda_{(2)} - \eta_{(1)} - \eta_{(2)} - \eta_{(3)} }{\Xi} u_{i}^{[1]} \right)\frac{\Delta b_{1i}}{c},
\end{equation}
where $\Delta b_{1i} := b_1 - b_i$ and $\Xi := \delta_2\left( \Lambda_{(2)} - \eta_{(1)} \right) - \Lambda_{(2)} + \eta_{(2)} + \eta_{(3)}$.

For a complete graph, the critical ratio for $\rho_{\textrm{C}} > 1/N$ is 
\begin{equation}
\left( \frac{b_1}{c} \right)^* = \frac{6\left( N-1 \right)}{\left( \delta_{2} + 2\right) \left( N - 2\right)} + \sum_{i=2}^{L}\left( \frac{\delta_2 + 1}{\delta_{2} + 2} u_{i}^{[2]} + \frac{1}{\delta_{2} + 2} u_{i}^{[1]} \right)\frac{\Delta b_{1i}}{c}.
\end{equation}
Letting $N \to \infty$, we have
\begin{equation}
\left( \frac{b_1}{c} \right)^* = \frac{6}{\left( \delta_{2} + 2\right)} + \sum_{i=2}^{L}\left( \frac{\delta_2 + 1}{\delta_{2} + 2} u_{i}^{[2]} + \frac{1}{\delta_{2} + 2} u_{i}^{[1]} \right)\frac{\Delta b_{1i}}{c}.
\end{equation}

For a ceiling fan graph, the critical ratio for $\rho_{\textrm{C}} > 1/N$ is 
\begin{equation}
\begin{split}
    \left( \frac{b_1}{c} \right)^* &= \frac{36 n \left( \substack{168 n^{3} + 178 n^{2}\\ - 137 n - 14} \right)}{ \delta_{2} \left( \substack{1288 n^{4} + 1181 n^{3}\\ - 1534 n^{2} - 338n\\ - 12 }\right) + \left( \substack{2114 n^{4} + 1636 n^{3}\\ - 2534 n^{2}  - 58n\\ + 12} \right)} + \sum_{i=2}^{L}\left( \frac{ \left( \delta_{2} + 1 \right) \left( \substack{1288 n^{4} + 1181 n^{3}\\ - 1534 n^{2} - 338 n\\ - 12} \right)}{\delta_{2} \left( \substack{1288 n^{4} + 1181 n^{3}\\ - 1534 n^{2} - 338 n\\ - 12} \right) + \left( \substack{2114 n^{4} + 1636 n^{3}\\ - 2534 n^{2} - 58 n\\ + 12} \right)}u_{i}^{[2]} \right.\\
    & \quad \left. + \frac{\left(\substack{826 n^{4} + 455 n^{3}\\ - 1000 n^{2} + 280 n\\ + 24} \right)}{ \delta_{2} \left( \substack{1288 n^{4} + 1181 n^{3}\\ - 1534 n^{2} - 338 n\\ - 12} \right) + \left( \substack{2114 n^{4} + 1636 n^{3}\\ - 2534 n^{2} - 58 n\\ + 12} \right)}u_{i}^{[1]} \right)\frac{\Delta b_{1i}}{c}.    
\end{split}
\end{equation}
Letting $n \to \infty$, we have
\begin{equation}
\left( \frac{b_1}{c} \right)^* = \frac{432}{92\delta_{2} + 151} + \sum_{i=2}^{L}\left( \frac{92\delta_{2} + 92}{92\delta_{2} + 151} u_{i}^{[2]} + \frac{59}{92\delta_{2} + 151} u_{i}^{[1]} \right)\frac{\Delta b_{1i}}{c}.
\end{equation}

For a conjoined star graph, the critical ratio for $\rho_{\textrm{C}} > 1/N$ is 
\begin{equation}
\begin{split}
    \left( \frac{b_1}{c} \right)^* &= \frac{n^{2} \left( \substack{1680 n^{5} + 5976 n^{4}\\ + 2760 n^{3} - 3528 n^{2}\\ - 792 n + 144 } \right)}{ \delta_{2} \left( \substack{532 n^{7} + 1576 n^{6}\\ + 281 n^{5} - 1022 n^{4}\\ + 296 n^{3} + 12 n^{2}\\ - 429 n + 54} \right) + \left( \substack{644 n^{7} + 2372 n^{6}\\ + 727 n^{5} - 1966 n^{4}\\ + 4 n^{3} + 204 n^{2}\\ - 183 n + 18} \right) } + \sum_{i=2}^{L}\left( \frac{ \left( \delta_{2} + 1 \right) \left( \substack{532 n^{7} + 1576 n^{6}\\ + 281 n^{5} - 1022 n^{4}\\ + 296 n^{3} + 12 n^{2}\\ - 429 n + 54} \right)}{\delta_{2} \left( \substack{532 n^{7} + 1576 n^{6}\\ + 281 n^{5} - 1022 n^{4}\\ + 296 n^{3} + 12 n^{2}\\ - 429 n + 54} \right) + \left( \substack{644 n^{7} + 2372 n^{6}\\ + 727 n^{5} - 1966 n^{4}\\ + 4 n^{3} + 204 n^{2}\\ - 183 n + 18} \right)}u_{i}^{[2]} \right. \\
    & \quad \left. + \frac{\left(\substack{ 112 n^{7} + 796 n^{6}\\ + 446 n^{5} - 944 n^{4}\\ - 292 n^{3} + 192 n^{2}\\ + 246 n - 36} \right)}{ \delta_{2} \left( \substack{532 n^{7} + 1576 n^{6}\\ + 281 n^{5} - 1022 n^{4}\\ + 296 n^{3} + 12 n^{2}\\ - 429 n + 54} \right) + \left( \substack{644 n^{7} + 2372 n^{6}\\ + 727 n^{5} - 1966 n^{4}\\ + 4 n^{3} + 204 n^{2}\\ - 183 n + 18} \right)} u_{i}^{[1]} \right)\frac{\Delta b_{1i}}{c}.
\end{split}
\end{equation}
Letting $n \to \infty$, we have
\begin{equation}
\left( \frac{b_1}{c} \right)^* = \frac{60}{19\delta_{2} + 23} + \sum_{i=2}^{L}\left( \frac{19\delta_{2} + 19}{19\delta_{2} + 23} u_{i}^{[2]} + \frac{4}{19\delta_{2} + 23} u_{i}^{[1]} \right)\frac{\Delta b_{1i}}{c}.
\end{equation}

Moreover, it is reasonable to analyze the case in which transitions can lead to the change of $\delta_2$, denoted by $\delta^i_2$ for $i \in \{ 1,2,\dots,L \}$. Consequently, the ``effective'' payoff structure can be described as
\begin{equation}
    \begin{pNiceMatrix}[first-row,first-col]
        & \textrm{C} & \textrm{D} \\
    \textrm{C} & -c + \sum_{i=1}^{L} u_{i}^{[2]} \frac{(\delta^{i}_{2} + 1)b}{2} & -c + \frac{b}{2} \\
    \textrm{D} & \frac{b}{2}  & 0 
    \end{pNiceMatrix}.
\end{equation}

According to the payoff structure, the coefficients of its payoff function are
\begin{equation}
\begin{split}
    B &= \frac{b}{2}, \\
    C &= \frac{b}{2} - c, \\
    D &= \sum_{i=1}^{L} u_{i}^{[2]} \frac{(\delta^{i}_{2} + 1)b}{2} - b.
\end{split}
\end{equation}
We immediately see that the expression of the critical ratio for $\rho_{\textrm{C}} > 1/N$ is
\begin{equation}
    \left( \delta^1_2 \right)^*= \frac{\Lambda_{(2)} - \eta_{(2)} - \eta_{(3)}}{\Lambda_{(2)} - \eta_{(1)}} + \left( \frac{\eta_{(2)}}{\Lambda_{(2)} - \eta_{(1)}} \right) \frac{2c}{b} + \sum_{i=2}^{L} u_{i}^{[2]} \Delta \delta^{1i}_{2}.    
\end{equation}

Generally, we can make the assumption that game transition can influence both the variance of $\delta_2$ and $b$. We can yield the corresponding expression with slight modifications.

\subsection{Transitions among Snowdrift Game}
Now let us move on to the discussion of the snowdrift game. For the snowdrift game with the payoff matrix
\begin{equation}
    \begin{pNiceMatrix}[first-row,first-col]
        & \textrm{C} & \textrm{D} \\
    \textrm{C} & b-\frac{c}{2} & b-c \\
    \textrm{D} & b & 0 
    \end{pNiceMatrix},
\end{equation}
substituting $B = b-c$, $C = b$ and $D = -b + c/2$ into Eq.~S\ref{fixation probability} gives 
\begin{equation}
\left( \frac{b}{c} \right)^* = \frac{ \Lambda_{(2)} - \eta_{(1)} - 2\eta_{(2)} }{2 \left(\Lambda_{(2)} - \eta_{(2)} - \eta_{(3)} \right)}.
\end{equation}

Similarly, the ``effective'' payoff structure of the snowdrift game can be described as 
\begin{equation}
    \begin{pNiceMatrix}[first-row,first-col]
        & \textrm{C} & \textrm{D} \\
    \textrm{C} & \sum_{i=1}^{L} u_{i}^{[2]} b_i - \frac{c}{2} & \sum_{i=1}^{L} u_{i}^{[1]} b_i - c \\
    \textrm{D} & \sum_{i=1}^{L} u_{i}^{[1]} b_i & 0 
    \end{pNiceMatrix}.
\end{equation}
According to the payoff structure, the coefficients of its payoff function are 
\begin{equation}
\begin{split}
    B &= \sum_{i=1}^{L} u_{i}^{[1]} b_i, \\
    C &= \sum_{i=1}^{L} u_{i}^{[1]} b_i - c, \\
    D &= \sum_{i=1}^{L} \left( u_{i}^{[2]} - 2u_{i}^{[1]} \right) b_i + \frac{c}{2}.
\end{split}
\end{equation}

As a consequence, the critical ratio for $\rho_{\textrm{C}} > 1/N$ is given by 
\begin{equation}
\left( \frac{b_1}{c} \right)^{*} = \frac{ \Lambda_{(2)} - \eta_{(1)} - 2\eta_{(2)} }{2 \left(\Lambda_{(2)} - \eta_{(2)} - \eta_{(3)} \right)} + \sum_{i=2}^{L} \left( \frac{ \Lambda_{(2)} - \eta_{(1)} }{\eta_{(2)} + \eta_{(3)} - \Lambda_{(2)}} u_{i}^{[2]} - \frac{ 2\Lambda_{(2)} - \eta_{(1)} - \eta_{(2)} - \eta_{(3)} }{\eta_{(2)} + \eta_{(3)} - \Lambda_{(2)}} u_{i}^{[1]}  \right) \frac{\Delta b_{1i}}{c}.    
\end{equation}

For a complete graph, the critical ratio for $\rho_{\textrm{C}} > 1/N$ is 
\begin{equation}
\left( \frac{b_1}{c} \right)^{*} = \frac{5 N - 4}{4 \left(N - 2\right)} +  \sum_{i=2}^{L} \left( \frac{1}{2} u_{i}^{[2]} + \frac{1}{2} u_{i}^{[1]} \right) \frac{\Delta b_{1i}}{c}.
\end{equation}
Letting $N \to \infty$, we have
\begin{equation}
\left( \frac{b_1}{c} \right)^{*} = \frac{5}{4} +  \sum_{i=2}^{L} \left( \frac{1}{2} u_{i}^{[2]} + \frac{1}{2} u_{i}^{[1]} \right) \frac{\Delta b_{1i}}{c}.
\end{equation}

For a ceiling fan graph, the critical ratio for $\rho_{\textrm{C}} > 1/N$ is 
\begin{equation}
\left( \frac{b_1}{c} \right)^{*} = \frac{\left( \substack{ 4760 n^{4} + 5227 n^{3}\\ - 3398 n^{2} - 166 n\\ + 12 } \right)}{4 \left( \substack{ 1057 n^{4} + 818 n^{3}\\ - 1267 n^{2} - 29 n\\ + 6 } \right)} + \sum_{i=2}^{L} \left( \frac{ \left( \substack{ 1288 n^{4} + 1181 n^{3}\\ - 1534 n^{2} - 338 n\\ - 12 } \right) }{2 \left( \substack{ 1057 n^{4} + 818 n^{3}\\ - 1267 n^{2} - 29 n\\ + 6 } \right)} u_{i}^{[2]}  + \frac{ \left( \substack{ 826 n^{4} + 455 n^{3}\\ - 1000 n^{2} + 280 n\\ + 24 } \right)}{2 \left( \substack{ 1057 n^{4} + 818 n^{3}\\ - 1267 n^{2} - 29 n\\ + 6 } \right) } u_{i}^{[1]} \right) \frac{\Delta b_{1i}}{c}
\end{equation}
Letting $n \to \infty$, we have 
\begin{equation}
\left( \frac{b_1}{c} \right)^{*} = \frac{170}{151}  + \sum_{i=2}^{L} \left( \frac{92}{151} u_{i}^{[2]} + \frac{59}{151} u_{i}^{[1]} \right) \frac{\Delta b_{1i}}{c}.
\end{equation}

For a conjoined star graph, the critical ratio for $\rho_{\textrm{C}} > 1/N$ is 
\begin{equation}
\left( \frac{b_1}{c} \right)^{*} = \frac{ \left( \substack{ 1148 n^{7} + 4400 n^{6} + 2479 n^{5}\\ - 2506 n^{4} - 1088 n^{3} + 132 n^{2}\\ + 429 n - 54 }\right)}{2 \left( \substack{ 644 n^{7} + 2372 n^{6} + 727 n^{5}\\ - 1966 n^{4} + 4 n^{3} + 204 n^{2}\\ - 183 n + 18 } \right)} + \sum_{i=2}^{L} \left( \frac{ \left( \substack{ 532 n^{7} + 1576 n^{6} + 281 n^{5}\\ - 1022 n^{4} + 296 n^{3} + 12 n^{2}\\ - 429 n + 54 } \right)}{ \left( \substack{ 644 n^{7} + 2372 n^{6} + 727 n^{5}\\ - 1966 n^{4} + 4 n^{3} + 204 n^{2}\\ - 183 n + 18 } \right) } u_{i}^{[2]} + \frac{2 \left( \substack{ 56 n^{7} + 398 n^{6} + 223 n^{5}\\ - 472 n^{4} - 146 n^{3} + 96 n^{2}\\ + 123 n - 18 } \right)}{ \left( \substack{ 644 n^{7} + 2372 n^{6} + 727 n^{5}\\ - 1966 n^{4} + 4 n^{3} + 204 n^{2}\\ - 183 n + 18} \right)} u_{i}^{[1]} \right) \frac{\Delta b_{1i}}{c}.        
\end{equation}
Letting $n \to \infty$, we have 
\begin{equation}
\left( \frac{b_1}{c} \right)^{*} = \frac{41}{46} + \sum_{i=2}^{L} \left( \frac{19}{23} u_{i}^{[2]} + \frac{4}{23} u_{i}^{[1]} \right) \frac{\Delta b_{1i}}{c}.
\end{equation}

\subsection{Homogeneous Game Transitions}
We consider exogenous game transitions among homogeneous games. Under the assumption that the probability of game $i$ is $p_i$ (see Main Text), indicating $\mathbf{P}^{[0]} = \mathbf{P}^{[1]} = \mathbf{P}^{[2]}$, therefore, for the donation game, its effective payoff structure is 
\begin{equation}
    \begin{pNiceMatrix}[first-row,first-col]
        & \textrm{C} & \textrm{D} \\
    \textrm{C} & \sum_{i=1}^{L} p_{i} b_i-c &   -c \\
    \textrm{D} & \sum_{i=1}^{L} p_{i} b_i & 0 
    \end{pNiceMatrix}.
\end{equation}
According to Eq.~S\ref{general result}, we obtain 
\begin{equation}
    \left( \frac{b_1}{c} \right)^{*}_{\text{DG}} = \frac{\eta_{(2)}}{\eta_{(3)} - \eta_{(1)}} + \sum_{i=2}^{L} p_i \frac{\Delta b_{1i}}{c}\,,
    \label{exo ratio DG}
\end{equation}
Similarly, we can derive the corresponding results for the public goods game and snowdrift game; they are
\begin{equation}
    \left( \frac{b_1}{c} \right)^{*}_{\text{PGG}} = \frac{2\eta_{(2)}}{\Xi} + \sum_{i=2}^{L} p_i \frac{\Delta b_{1i}}{c}\,,
    \label{exo ratio PGG}
\end{equation}
and
\begin{equation}
   \left( \frac{b_1}{c} \right)^{*}_{\text{SG}} = \frac{ \Lambda_{(2)} - \eta_{(1)} - 2\eta_{(2)} }{2 \left(\Lambda_{(2)} - \eta_{(2)} - \eta_{(3)} \right)} + \sum_{i=2}^{L} p_i \frac{\Delta b_{1i}}{c}\,,
    \label{exo ratio SG}
\end{equation}
respectively.

\subsection{Heterogeneous Game Transitions}
The assumption that game transitions are homogeneous is made in our above study. The homogeneity refers to the transitions within one fixed cooperative game, but with respect to different game intensities. Therefore, we introduce game transitions with heterogeneity to model the changing environments based on distinct social contexts. 

We denote the situation in which mutual cooperation leads to DG and brings about SG, otherwise, as DS, of which the effective payoff configuration is 
\begin{equation}
    \begin{pNiceMatrix}[first-row,first-col]
        & \textrm{C} & \textrm{D} \\
    \textrm{C} & b-c & b-c \\
    \textrm{D} & b & 0 
    \end{pNiceMatrix}.
\end{equation}
Then, the critical ratio for such a case is given by 
\begin{equation}
\left(\frac{b}{c} \right)^{*} = \frac{\eta_{(2)}}{\eta_{(2)} + \eta_{(3)} - \Lambda_{(2)}}.
\end{equation}

For complete graphs, it becomes 
\begin{equation}
    \left(\frac{b}{c} \right)^{*} = \frac{3 \left(N - 1\right)}{2 \left(N - 2\right)}.
\end{equation}

For ceiling fan graphs, it becomes
\begin{equation}
\left(\frac{b}{c} \right)^{*} = \frac{9 n \left(168 n^{3} + 178 n^{2} - 137 n - 14\right)}{1057 n^{4} + 818 n^{3} - 1267 n^{2} - 29 n + 6}.
\end{equation}

For conjoined star graphs, it becomes
\begin{equation}
\left(\frac{b}{c} \right)^{*} = \frac{n^{2} \left(840 n^{5} + 2988 n^{4} + 1380 n^{3} - 1764 n^{2} - 396 n + 72\right)}{644 n^{7} + 2372 n^{6} + 727 n^{5} - 1966 n^{4} + 4 n^{3} + 204 n^{2} - 183 n + 18}.    
\end{equation}

We denote the situation in which mutual cooperation leads to SG and brings about DG, otherwise, as SG, of which the effective payoff structure is 
\begin{equation}
    \begin{pNiceMatrix}[first-row,first-col]
        & \textrm{C} & \textrm{D} \\
    \textrm{C} & b-\frac{c}{2} & -c \\
    \textrm{D} & b & 0 
    \end{pNiceMatrix}.
\end{equation}
Then, the critical ratio for such a case is given by 
\begin{equation}
\left(\frac{b}{c} \right)^{*} = \frac{2\eta_{(2)} + \eta_{(1)} - \Lambda_{(2)}}{2\eta_{(3)} - 2\eta_{(2)}}.
\end{equation}

For complete graphs, it becomes
\begin{equation}
    \left(\frac{b}{c} \right)^{*} = \frac{5 N^{2} - 14 N + 8}{6}.
\end{equation}

For ceiling fan graphs, it becomes
\begin{equation}
\left(\frac{b}{c} \right)^{*} = \frac{4760 n^{4} + 5227 n^{3} - 3398 n^{2} - 166 n + 12}{18 n \left(98 n^{3} + 133 n^{2} - 32 n - 4\right)}.
\end{equation}

For conjoined star graphs, it becomes
\begin{equation}
\left(\frac{b}{c} \right)^{*} = \frac{1148 n^{8} + 3252 n^{7} - 1921 n^{6} - 4985 n^{5} + 1418 n^{4} + 1220 n^{3} + 297 n^{2} - 483 n + 54}{12 \left(84 n^{8} + 240 n^{7} - 51 n^{6} - 151 n^{5} + 245 n^{4} - 83 n^{3} - 126 n^{2} + 114 n - 12\right)}.   
\end{equation}

Now we turn to exogenous transitions with heterogeneity. We analyze the following simple case, where DG and SG uniformly appear with probabilities $p_1$ and $p_2$ ($p_1 + p_2 = 1$). Then we see that 
\begin{equation}
    \left(\frac{b}{c} \right)^{*}_{\text{DS}} = \frac{p_2\eta_{(1)} + 2\eta_{(2)} - p_2\Lambda_{(2)}}{-2p_1\eta_{(1)} + 2p_2\eta_{(2)} + 2\eta_{(3)} - 2p_2\Lambda_{(2)}}.
\end{equation}
    
We start from complete networks with size $N$, of which the critical ratio is 
\begin{equation}
    \left(\frac{b}{c} \right)^{*}_{\text{DS}} = \frac{ \left( p_1+5 \right)N - 2 p_1 - 4}{4 \left( 1-p_1 \right) N + 2 \left( p_1-4 \right)}.
\end{equation}

We then focus on star graphs with $n$ leaves, which indicate the following recurrence relations
\begin{equation}
\begin{split}
\eta_{r_1l_1} &= \frac{1}{2} + \frac{1}{2} \left( \frac{n-1}{n} \eta_{l_1l_2} \right),\\
\eta_{l_1l_2} &= \frac{1}{2} + \eta_{r_1l_1},\\
\eta_{r_1l_1l_2} &= \frac{1}{3} + \frac{1}{3} \left( \frac{n-2}{n}\eta_{l_1l_2l_3} + \frac{2}{n}\eta_{l_1l_2} + 2\eta_{r_1l_1} \right),\\
\eta_{l_1l_2l_3} &= \frac{1}{3} + \eta_{r_1l_1l_2},\\
\end{split}
\end{equation}
which gives 
\begin{equation}
\begin{split}
\eta_{r_1l_1} &= \frac{3n-1}{2\left( n+1 \right)},\\
\eta_{l_1l_2} &= \frac{2n}{n+1},\\
\eta_{r_1l_1l_2} &= \frac{13n-2}{6\left( n+1 \right)},\\
\eta_{l_1l_2l_3} &= \frac{5n}{2\left( n+1 \right)},\\
\eta_{(1)} &= \frac{3n-1}{2 \left( n+1\right)},\\
\eta_{(2)} &= \frac{n-1}{n+1},\\
\eta_{(3)} &= \frac{3n-1}{2 \left( n+1\right)},\\
\Lambda_{(2)} &= \frac{22n^2-9n-1}{12n\left( n+1 \right)}.\\
\end{split}
\end{equation}
Following the case discussed in the main text, we have the following critical ratio  
\begin{equation}
\left(\frac{b}{c} \right)^{*}_{\text{DS}} = \frac{ \left( 4p_1+20 \right) n + p_1 - 1}{2 \left( \left( 8-8p_1 \right) n + p_1 - 1\right)}.
\end{equation}
for star graphs, where $0 \leq p_1 < 1$. Letting $n \to \infty$, we have
\begin{equation}
\left(\frac{b}{c} \right)^{*}_{\text{DS}} = \frac{p_1+5}{4-4p_1}.
\end{equation}

\subsection{Effects of Game Transitions on Costs}
The effective payoff structure for the donation game is
\begin{equation}
    \begin{pNiceMatrix}[first-row,first-col]
        & \textrm{C} & \textrm{D} \\
    \textrm{C} & b-c_1 & -c_2 \\
    \textrm{D} & b & 0 
    \end{pNiceMatrix}.
\end{equation}
After some straightforward derivations, we see that
\begin{equation}
    \left( c_1 \right)^{*} = \frac{\eta_{(1)}+\eta_{(2)}-\Lambda_{(2)}}{\eta_{(2)}}\Delta c + \frac{\eta_{(3)}-\eta_{(1)}}{\eta_{(2)}}b\,.
\end{equation}

The effective payoff structure for the snowdrift game is
\begin{equation}
    \begin{pNiceMatrix}[first-row,first-col]
        & \textrm{C} & \textrm{D} \\
    \textrm{C} & b-\frac{c_1}{2} & b-c_2 \\
    \textrm{D} & b & 0 
    \end{pNiceMatrix}.
\end{equation}
As before, we arrive at
\begin{equation}
    \left( c_1 \right)^{*} = \frac{2\left( \eta_{(1)}+\eta_{(2)}-\Lambda_{(2)} \right)}{\eta_{(1)}+2\eta_{(2)}-\Lambda_{(2)}}\Delta c + \frac{2\left( \eta_{(2)} + \eta_{(3)}-\Lambda_{(2)} \right)}{\eta_{(1)}+2\eta_{(2)}-\Lambda_{(2)}}b\,.
\end{equation}

\section{Relative Measure of Success}
Here, we consider cooperation is favored against defection in the sense of $\rho_{\textrm{C}} > \rho_{\textrm{D}}$. Derivations can be made with slight modifications to the case of $\rho_{\textrm{C}} > 1/N$ (details omitted).
\subsection{Complete Graphs}
For a complete graph of size $N$, we have
\begin{equation}
    \tau_{ij} = N-1, \, \tau_{ijk} = \frac{3(N-1)}{2},
\end{equation}
and 
\begin{equation}
\begin{split}
    \tau_{(1)} &= N-1, \\ 
    \tau_{(2)} &= N-2, \\ 
    \tau_{(3)} &= \frac{N^2-3N+3}{N-1}, \\
    \Gamma_{(2)} &= \frac{3N^2-8N+6}{2(N-1)}.
\end{split}
\end{equation}

Therefore, the critical ratio for $\rho_{\textrm{C}} >\rho_{\textrm{D}}$ is given by
\begin{equation}
\left( \frac{b_1}{c}\right)^* = -N + 1 + 
\sum_{i=2}^{L} \left(-\frac{N^2 -2N +2}{2\left(N - 2\right)} u^{[2]}_{i} + \frac{N}{2} u^{[1]}_{i} \right) \frac{\Delta b_{1i}}{c}.
\end{equation}

\subsection{Ceiling Fan Graphs}
For a ceiling fan graph with $2n$ leaves, we have
\begin{equation}
\begin{split}
\tau_{r l_{1}} &= \frac{2\left(8n-3\right)}{2n+3}, \\ 
\tau_{l^{+}_{1} l^{-}_{1}} &= \frac{10n}{2n+3}, \\
\tau_{l_{1} l_{2}} &= \frac{20n}{2n+3}, \\
\tau_{r l^{+}_{1} l^{-}_{1}} &= \frac{3 \left(7 n - 2\right)}{2 n + 3}, \\
\tau_{r l_{1} l_{2}} &= \frac{2 \left(13 n - 3\right)}{2 n + 3}, \\
\tau_{l^{+}_{1} l^{-}_{1} l_{2}} &= \frac{25 n}{2 n + 3}, \\
\tau_{l_{1} l_{2} l_{3}} &= \frac{30 n}{2 n + 3}.
\end{split}
\end{equation}

Then, we have
\begin{equation}
\begin{split}
\tau_{(1)} &= \frac{2\left(21n-6\right)}{3\left(2n+3\right)}, \\
\tau_{(2)} &= \frac{36n-21}{3\left(2n+3\right)}, \\
\tau_{(3)} &= \frac{93n-48}{6\left(2n+3\right)}, \\ 
\Gamma_{(2)} &= \frac{83n-38}{4\left(2n+3\right)}.
\end{split}
\end{equation}

Therefore, the critical ratio for $\rho_{\textrm{C}} >\rho_{\textrm{D}}$ is given by
\begin{equation}
\left( \frac{b_1}{c}\right)^* = \frac{2\left(12n-7\right)}{3n-8} + 
\sum_{i=2}^{L} \left( \frac{27 n - 22}{2 \left(3 n - 8\right)} u^{[2]}_{i} - 
\frac{3 \left(7 n - 2\right)}{2 \left(3 n - 8\right)} u^{[1]}_{i} \right) \frac{\Delta b_{1i}}{c}.
\end{equation}

\subsection{Conjoining Two Star Graphs}
For a network by conjoining two identical star graphs of size $n$, the following equalities hold
\begin{equation}
\begin{split}
    \tau_{r_1r_2} &= \frac{4n^3+8n^2-11n+4}{n\left(2n+3\right)}, \\
    \tau_{r_1l_1} &= \frac{10n-5}{2n+3}, \\
    \tau_{r_1l_2} &= \frac{4n^2+10n-4}{2n+3}, \\
    \tau_{l^+_1l^-_1} &= \frac{12n-2}{2n+3}, \\
    \tau_{l_1l_2} &= \frac{4n^2+12n-1}{2n+3}, \\
    \tau_{r_1l^+_1l^-_1} &= \frac{16 n - 6}{2 n + 3}, \\
    \tau_{r_1r_2l_1} &= \frac{4 n^{3} + 14 n^{2} - 10 n + 2}{2 n^{2} + 3 n}, \\ 
    \tau_{r_1l_1l_2} &= \frac{4 n^{2} + 16 n - 5}{2 n + 3}, \\
    \tau_{r_1l^+_2l^-_2} &= \frac{4 n^{2} + 16 n - 5}{2 n + 3}, \\
    \tau_{l^+_1l^-_1l_2} &= \frac{4 n^{2} + 18 n - 2}{2 n + 3}, \\
    \tau_{l^+_1l^-_1l^*_1} &= \frac{18 n - 3}{2 n + 3}.
\end{split}
\end{equation}
As a consequence, we have
\begin{equation}
\begin{split}
\tau_{(1)} &= \frac{24 n^{3} - 22 n^{2} - n + 4}{n \left(4 n^{2} + 4 n - 3\right)}, \\
\tau_{(2)} &= \frac{2 \left(10 n^{3} - 13 n^{2} + n + 2\right)}{n \left(4 n^{2} + 4 n - 3\right)}, \\
\tau_{(3)} &= \frac{32 n^{5} - 38 n^{4} + 5 n^{3} + 12 n^{2} - 10 n + 4}{n^{3} \left(4 n^{2} + 4 n - 3\right)}, \\ 
\Gamma_{(2)} &= \frac{38 n^{5} - 43 n^{4} + 3 n^{3} + 10 n^{2} - 5 n + 2}{n^{3} \left(4 n^{2} + 4 n - 3\right)}.   
\end{split}
\end{equation}

Therefore, the critical ratio for $\rho_{\textrm{C}} >\rho_{\textrm{D}}$ is given by
\begin{equation}
\left( \frac{b_1}{c} \right)^* = \frac{n^{2} \left(10 n^{2} - 3 n - 2\right)}{\left(n - 1\right) \left(4 n^{3} - n + 2\right)} + 
\sum_{i=2}^{L} \left( \frac{\left( \substack{ 14 n^{4} - 7 n^{3} - 3 n^{2}\\ + 3 n - 2 } \right)}{2 \left( \substack{ 4 n^{4} - 4 n^{3} - n^{2}\\ + 3 n - 2 }\right)} u^{[2]}_{i} - 
\frac{\left(\substack{6 n^{4} + n^{3} - n^{2}\\ - 3 n + 2} \right)}{2 \left( \substack{ 4 n^{4} - 4 n^{3} - n^{2}\\ + 3 n - 2 }\right)} u^{[1]}_{i} \right) \frac{\Delta b_{1i}}{c}.
\end{equation}

\section{Evolution of Cooperation with Mutation}
In the presence of mutation (mutation rate denoted by $u$), cooperation is favored if the mean cooperation frequency $\left<f_{\textrm{C}}\right>$ is greater than $1/2$. With the straightforward application of the framework proposed in Ref. \cite{mcavoy2022evaluating}, we can obtain the desired results. For that purpose, we briefly recall the relevant results. Eq.~$10$ in Ref.~\cite{mcavoy2022evaluating} gives
\begin{equation}
\left<f_{\textrm{C}}\right> = \frac{1}{2} + \frac{\delta}{u} \sum_{i,j,k=1}^{N} \pi^{\mathrm{mut}}_{i} m^{ji}_{k} \left( \frac{1-u}{2} \left< U_k|j \sim \textrm{C} \right>^{\circ} + \frac{u}{2} \left< U_k \right>^{\circ} - \frac{1}{2} \left< U_k|i  \sim \textrm{C} \right>^{\circ} \right) + O(\delta^2),
\end{equation}
where $\pi^{\mathrm{mut}}_{i}$ is the mutation-weighted reproductive value and $m^{ji}_{k}$ represents the marginal effect of $k$'s fitness on $i$ replacing $j$. For games with two strategies, it can be reduced to a simpler form, i.e.,
\begin{equation}
    \left<f_{\textrm{C}}\right> = \frac{1}{2} + \frac{\delta}{2} \left( K_1 \left( a_{11} - a_{22} \right) + K_2 \left( a_{12} - a_{21} \right) \right) + O(\delta^2), 
\end{equation}
where 
\begin{equation}
\begin{split}
    K_1 &= \frac{1}{2u} \sum_{i,j,k = 1}^{N} \pi^{\mathrm{mut}}_{i}m^{ji}_{k} \sum^{N}_{l=1} \Omega_{kl} \left( -\left( \phi_{ik} + \phi_{il} \right) + \left( 1-u \right) \left( \phi_{jk} + \phi_{jl} \right) + u \right) \\
    K_2 &= \frac{1}{2u} \sum_{i,j,k = 1}^{N} \pi^{\mathrm{mut}}_{i}m^{ji}_{k} \sum^{N}_{l=1} \Omega_{kl} \left( -\left( \phi_{ik} - \phi_{il} \right) + \left( 1-u \right) \left( \phi_{jk} - \phi_{jl} \right)\right)
\end{split}.
\end{equation}
$\phi_{ij}$ and $\Omega_{ij}$ measure the probability that $i$ and $j$ are identical-by-state and the interaction rule of the population respectively.

Substituting the related payoff configuration and specified population structure into the above equation, we can get the mean cooperation frequency for selection-mutation-game evolutionary dynamics. Therefore, the critical ratio for $\left<f_{\textrm{C}}\right> > 1/2$ with single donation game is 
\begin{equation}
\left( \frac{b}{c} \right)^* = \frac{K_1 + K_2}{K_1 - K_2},
\end{equation}
and that for the deterministic game transition is 
\begin{equation}
\left( \frac{b_1}{c} \right)^* = \frac{K_1 + K_2}{K_1 - K_2} - \frac{K_2}{K_1 - K_2} \cdot \frac{\Delta b}{c}\,.
\end{equation}
Analogously, the critical threshold for $\left<f_{\textrm{C}}\right> > 1/2$ with single snowdrift game is 
\begin{equation}
\left( \frac{b}{c} \right)^* = \frac{K_1 + 2K_2}{2K_1},
\end{equation}
and that for its deterministic game transition is 
\begin{equation}
\left( \frac{b_1}{c} \right)^* = \frac{K_1 + 2K_2}{2K_1}.
\end{equation}

Under death-Birth update the pairwise identity-by-state probabilities $\phi$ satisfy $\phi_{ii} = 1$ and for $i \neq j$,
\begin{equation}
\phi_{ij} = \frac{u}{2} + \frac{1-u}{2} \sum^{N}_{k=1}A_{ik}\phi_{kj} + \frac{1-u}{2} \sum^{N}_{k=1}A_{jk}\phi_{ik}.
\end{equation}
The marginal effects of selection is calculated by $m^{ij}_{i} = A_{ji}\left( 1-A_{ji} \right) / N$ and $m^{ij}_{k} = -A_{ji}A_{jk} / N$ for $i \neq k$, where $A_{ij}=w_{ij}/w_i$. Finally, the mutation-weighted reproductive value is given by $\pi^{\mathrm{mut}}_{i} = \sum^{N}_{j=1} u\left( \left( I - \left( 1-u \right)A \right)^{-1} \right)_{ji}$\,. 

\section{Pairwise-comparison Updating}
We consider another strategy update rule, pairwise-comparison (PC) updating. Under PC, a randomly chosen individual (denoted $i$) chooses one of its neighbors at random (denoted $j$) and imitates its strategy with probability $\left(1+\exp\left\{-\delta\left(f_{j}-f_{i}\right)\right\}\right)^{-1}$, where $f_i$ (resp. $f_j$) is the payoff of $i$ (resp. $j$). According to the rule of PC updating, as before, we can obtain

\begin{align}
c_{k}^{ij} &= 
\begin{cases}
\frac{p^{(1)}_{ji}}{4N} \left(C \left(p^{(0)}_{ik} - p^{(0)}_{jk}\right) + B \left( p^{(1)}_{ik} - p^{(1)}_{jk} \right) \right) & i\neq j \\
& \\
\frac{1}{4N} \left(C \left(p^{(0)}_{jk} - p^{(1)}_{jk}\right) + B \left( p^{(1)}_{jk} - p^{(2)}_{jk} \right) \right) & i = j 
\end{cases},
\end{align}
and
\begin{align}
c_{kl}^{ij} &= 
\begin{cases}
D\frac{p^{(1)}_{ji}}{4N}\left( p^{(0)}_{ik}p^{(1)}_{il} - p^{(0)}_{jk}p^{(1)}_{jl} + p^{(0)}_{il}p^{(1)}_{ik} - p^{(0)}_{jl}p^{(1)}_{jk} \right)  & i\neq j \\
& \\
D\frac{1}{4N} \left( p^{(1)}_{jk}p^{(0)}_{jl} - p^{(1)}_{jl}p^{(1)}_{lk} + p^{(1)}_{jl}p^{(0)}_{jk} - p^{(1)}_{jk}p^{(1)}_{kl} \right)  & i = j 
\end{cases}.
\end{align}
Then, as before, we arrive at
\begin{equation}
\rho_{\textrm{C}}(\bm{\xi}) = \widehat{\xi} + \frac{\delta}{N} \left[ C \eta^{\bm{\xi}}_{(1)} + B\left( \eta^{\bm{\xi}}_{(2)} - \eta^{\bm{\xi}}_{(1)} \right) + D \left(\Lambda^{\bm{\xi}}_{(1)} - \eta^{\bm{\xi}}_{(1)} \right)\right] + O(\delta^2),  
\label{fixation probability PC}
\end{equation}
where $B=a_{21}-a_{22}$, $C=a_{12}-a_{22}$, and $D=a_{11}-a_{12}-a_{21}+a_{22}$.

For the deterministic game transition in the donation game, we have 
\begin{equation}
\left( \frac{b_1}{c} \right)^* = \frac{\eta_{(1)}}{\eta_{(2)}-\eta_{(1)}} - \frac{\Lambda_{(1)} - \eta_{(2)}}{\eta_{(2)}-\eta_{(1)}} \cdot \frac{\Delta b}{c},
\end{equation}
and its equivalent form
\begin{equation}
\left( \frac{\Delta b}{c} \right)^* = \frac{\eta_{(1)}}{\Lambda_{(1)}-\eta_{(1)}} - \frac{\eta_{(2)} - \eta_{(1)}}{\Lambda_{(1)}-\eta_{(1)}} \cdot \frac{b_2}{c}.
\end{equation}

For the deterministic game transition in the snowdrift game, we have 
\begin{equation}
\left( \frac{b_1}{c} \right)^* = \frac{3\eta_{(1)}-\Lambda_{(1)}}{2\eta_{(1)}+2\eta_{(2)}-2\Lambda_{(1)}} + \frac{2\eta_{(1)}+\eta_{(2)}-2\Lambda_{(1)}}{\eta_{(1)}+\eta_{(2)}-\Lambda_{(1)}} \cdot \frac{\Delta b}{c},
\end{equation}
and its equivalent form
\begin{equation}
\left( \frac{\Delta b}{c} \right)^* = \frac{3\eta_{(1)}-\Lambda_{(1)}}{2\Lambda_{(1)}-2\eta_{(1)}} - \frac{\eta_{(1)} + \eta_{(2)} - \Lambda_{(1)}}{\Lambda_{(1)}-\eta_{(1)}} \cdot \frac{b_2}{c}.
\end{equation}

\section{More Discussions}
Recall that for complete graphs, the following equalities hold 
\begin{equation}
\begin{split}
    \tau_{(1)} &= N-1, \\ 
    \tau_{(2)} &= N-2, \\
    \tau_{(3)} &= \frac{N^2-3N+3}{N-1} \\
    \Gamma_{(0)} &= N-1, \\
    \Gamma_{(1)} &= \frac{3N-4}{2}, \\
    \Gamma_{(2)} &= \frac{3 N^{2} - 8 N + 6}{2 \left(N - 1\right)}. 
\end{split}
\end{equation}
For ceiling fan graphs, we have the subsequent equalities 
\begin{equation}
\begin{split}
    \tau_{(1)} &= \frac{2\left(21n-6\right)}{3\left(2n+3\right)}, \\
    \tau_{(2)} &= \frac{36n-21}{3\left(2n+3\right)}, \\
    \tau_{(3)} &= \frac{93n-48}{6\left(2n+3\right)}, \\ 
    \Gamma_{(0)} &= \frac{2\left(21n-6\right)}{3\left(2n+3\right)}, \\
    \Gamma_{(1)} &= \frac{5 \left(8 n - 3\right)}{2 \left(2 n + 3\right)}, \\
    \Gamma_{(2)} &= \frac{83n-38}{4\left(2n+3\right)} .
\end{split}
\end{equation}
For conjoined star graphs, the following equalities hold
\begin{equation}
\begin{split}
\tau_{(1)} &= \frac{24 n^{3} - 22 n^{2} - n + 4}{n \left(4 n^{2} + 4 n - 3\right)}, \\
\tau_{(2)} &= \frac{2 \left(10 n^{3} - 13 n^{2} + n + 2\right)}{n \left(4 n^{2} + 4 n - 3\right)}, \\
\tau_{(3)} &= \frac{32 n^{5} - 38 n^{4} + 5 n^{3} + 12 n^{2} - 10 n + 4}{n^{3} \left(4 n^{2} + 4 n - 3\right)}, \\ 
\Gamma_{(0)} &= \frac{24 n^{3} - 22 n^{2} - n + 4}{n \left(4 n^{2} + 4 n - 3\right)}, \\
\Gamma_{(1)} &= \frac{34 n^{3} - 35 n^{2} + 6}{n \left(4 n^{2} + 4 n - 3\right)}, \\
\Gamma_{(2)} &= \frac{38 n^{5} - 43 n^{4} + 3 n^{3} + 10 n^{2} - 5 n + 2}{n^{3} \left(4 n^{2} + 4 n - 3\right)}.   
\end{split}
\end{equation}

Note that, for the discussed complete graphs, ceiling fan graphs, and conjoined star graphs, the following relations are established, i.e.,
\begin{equation}
\begin{split}
    \Gamma_{(0)} &= \tau_{(1)}, \\
    \Gamma_{(1)} &= \tau_{(1)} + \frac{\tau_{(2)}}{2}, \\
    \Gamma_{(2)} &= \frac{\tau_{(1)} + \tau_{(2)} + \tau_{(3)}}{2}.     
\end{split}
\end{equation}
Based on it, we infer that the following recurrence relation ($n \geq 0$)
\begin{equation}
    \Gamma_{(n+1)} = \Gamma_{(n)} + \frac{\tau_{(n+2)}-\tau_{(n)}}{2}
\end{equation}
hold for any connected graphs. 

By introducing the expected remeeting time $\tau^{+}_{ii} := 1 + \sum_{j \in \mathcal{G}} p^{(1)}_{ij}\tau_{ij}$, the recurrence relation for $\tau_{(n)}$ is then given by \cite{allen2017evolutionary}
\begin{equation}
    \tau_{(n+1)} = \tau_{(n)} + \sum_{i \in \mathcal{G}} \pi_i p^{(n)}_{ii} \tau^{+}_{ii} - 1.
\end{equation}
Therefore, the recurrence relation becomes
\begin{equation}
    \Gamma_{(n+1)} = \Gamma_{(n)} + \frac{\sum_{i \in \mathcal{G}} \pi_i \left( p^{(n)}_{ii} + p^{(n+1)}_{ii} \right) \tau^{+}_{ii}}{2} - 1.
\end{equation}

Now we turn to the case where $|I|=4$. We define $\Phi_{(n)} := \sum_{i_1,i_2,i_3,i_4 \in \mathcal{G}} \pi_{i_1} p^{(n)}_{i_1i_2}p^{(1)}_{i_2i_3} p^{(1)}_{i_3i_4} \tau_{i_1i_2i_3i_4}$ and consider the trivial topologies, i.e., the complete graphs. For $\forall i_1, i_2, i_3, i_4 \in \mathcal{G}$, we have 
\begin{equation}
\begin{split}
\tau_{i_1i_2i_3i_4} &= 1 + \frac{1}{4} \left( \sum_{y \in \mathcal{G}} p^{(1)}_{i_1y} \tau_{yi_2i_3i_4} + p^{(1)}_{i_2y} \tau_{i_1yi_3i_4} + p^{(1)}_{i_3y} \tau_{i_1i_2yi_4} + p^{(1)}_{i_4y} \tau_{i_1i_2i_3y}\right) \\
&= \frac{11}{2} + \frac{N-4}{N-1} \tau_{i_1i_2i_3i_4},
\end{split}
\end{equation}
which gives 
\begin{equation}
\tau_{i_1i_2i_3i_4} = \frac{11\left( N - 1 \right)}{6}.
\end{equation}

The equality $\Phi_{(0)} = \left( 3N-4 \right)/2$ follows immediately from the definition of $\Phi_{(n)}$. Based on it, we have
\begin{equation}
\Phi_{(1)} = \frac{11 N^{2} - 28 N + 18}{6 \left(N - 1\right)},
\end{equation}
while
\begin{equation}
\tau_{(1)} + \frac{\tau_{(2)}}{2} + \frac{\tau_{(3)}}{3} = \frac{11 N^{2} - 27 N + 18}{6 \left(N - 1\right)}.
\end{equation}
It shows that the following recurrence relation ($n \geq 0$)
\begin{equation}
    \Phi_{(n+1)} = \Phi_{(n)} + \frac{\tau_{(n+3)}-\tau_{(n)}}{3}
\end{equation}
or its equivalent form
\begin{equation}
\Phi_{(n+1)} = \Phi_{(n)} + \frac{\sum_{i \in \mathcal{G}} \pi_i \left( p^{(n)}_{ii} + p^{(n+1)}_{ii} + p^{(n+2)}_{ii}\right) \tau^{+}_{ii}}{3} - 1
\end{equation}
does not hold even in complete networks.

For the equality $\Gamma_{(2)} = (\tau_{(1)} + \tau_{(2)} + \tau_{(3)}) / 2$, we provide the following heuristic proof. For the deterministic game transition with the effective payoff structure
\begin{equation}
    \begin{pNiceMatrix}[first-row,first-col]
        & \textrm{C} & \textrm{D} \\
    \textrm{C} &  b_1-c &   -c \\
    \textrm{D} &  b_2 & 0 
    \end{pNiceMatrix},
\end{equation}
applying the Structure Coefficient Theorem \cite{tarnita2009strategy, allen2017evolutionary}, $\sigma a + b > c + \sigma d$, where
\begin{equation}
    \sigma = \frac{-\tau_{(1)} + \tau_{(2)} + \tau_{(3)}}{\tau_{(1)}+\tau_{(2)}-\tau_{(3)}},
\end{equation}
we immediately see that 
\begin{equation}
    \left( \frac{b_1}{c}\right)^* = \frac{\tau_{(2)}}{\tau_{(3)} - \tau_{(1)}} - \frac{\tau_{(1)} + \tau_{(2)} - \tau_{(3)}}{2\left( \tau_{(3)} - \tau_{(1)} \right)} \cdot \frac{\Delta b_{12}}{c}.
\end{equation}
Comparing it to the expression given before (Eq.~S\ref{end critical ratio}), we arrive at the desired equality $\Gamma_{(2)} = (\tau_{(1)} + \tau_{(2)} + \tau_{(3)}) / 2$. For the general case, we leave it for future investigation.

Therefore, the critical ratio for $\rho_{\textrm{C}} > \rho_{\textrm{D}}$ can be simplified as
\begin{equation}
\left( \frac{b_1}{c}\right)^* = \frac{\tau_{(2)}}{\tau_{(3)} - \tau_{(1)}} + 
\sum^{L}_{i=2}\left( \frac{-\tau_{(1)} + \tau_{(2)} + \tau_{(3)}}{2\left( \tau_{(3)} - \tau_{(1)} \right)}u^{[2]}_{i} - 
\frac{\tau_{(1)} + \tau_{(2)} - \tau_{(3)}}{2\left( \tau_{(3)} - \tau_{(1)} \right)}u^{[1]}_{i}\right) \frac{\Delta b_{1i}}{c}.
\end{equation}
Furthermore, for regular graphs with degree $k$ and size $N$, the critical ratio becomes
\begin{equation}
\left( \frac{b_1}{c}\right)^* = \frac{N - 2}{N/k - 2} + 
\sum^{L}_{i=2}\left( \frac{N + N/k - 4}{2\left( N/k - 2 \right)}u^{[2]}_{i} - 
\frac{N - N/k}{2\left( N/k - 2 \right)}u^{[1]}_{i}\right) \frac{\Delta b_{1i}}{c},
\end{equation}
due to $\tau_{(1)} = N-1$, $\tau_{(2)} = N-2$ and $\tau_{(3)} = N+N/k-3$. Letting $N \rightarrow \infty$, we obtain the result in Ref. \cite{su2019evolutionary}, i.e., 
\begin{equation}
\left( \frac{b_1}{c} \right)^* = k + \sum^{L}_{i=2}\left( \frac{k+1}{2}u^{[2]}_{i} - \frac{k - 1}{2}u^{[1]}_{i}\right) \frac{\Delta b_{1i}}{c}.
\end{equation}

For the public goods game, we see that 
\begin{equation}
    \left( \frac{b_1}{c} \right)^* = \frac{4\tau_{(2)}}{\Xi} + \sum_{i=2}^{L} u_{i}^{[2]} \frac{\Delta b_{1i}}{c},    
\end{equation}
where $\Xi := \left( \delta_2 + 1 \right) \left( -\tau_{(1)} + \tau_{(2)} + \tau_{(3)} \right)$.

For the snowdrift game, it becomes
\begin{equation}
    \left( \frac{b_1}{c} \right)^{*} = \frac{ \tau_{(1)} + 3\tau_{(2)} - \tau_{(3)}}{2 \left( -\tau_{(1)} + \tau_{(2)} + \tau_{(3)} \right)} + \sum_{i=2}^{L} u_{i}^{[2]} \frac{\Delta b_{1i}}{c}.    
\end{equation}

\clearpage
\renewcommand{\figurename}{Supplementary Figure}

\begin{figure}
\centering
\includegraphics[width=\textwidth]{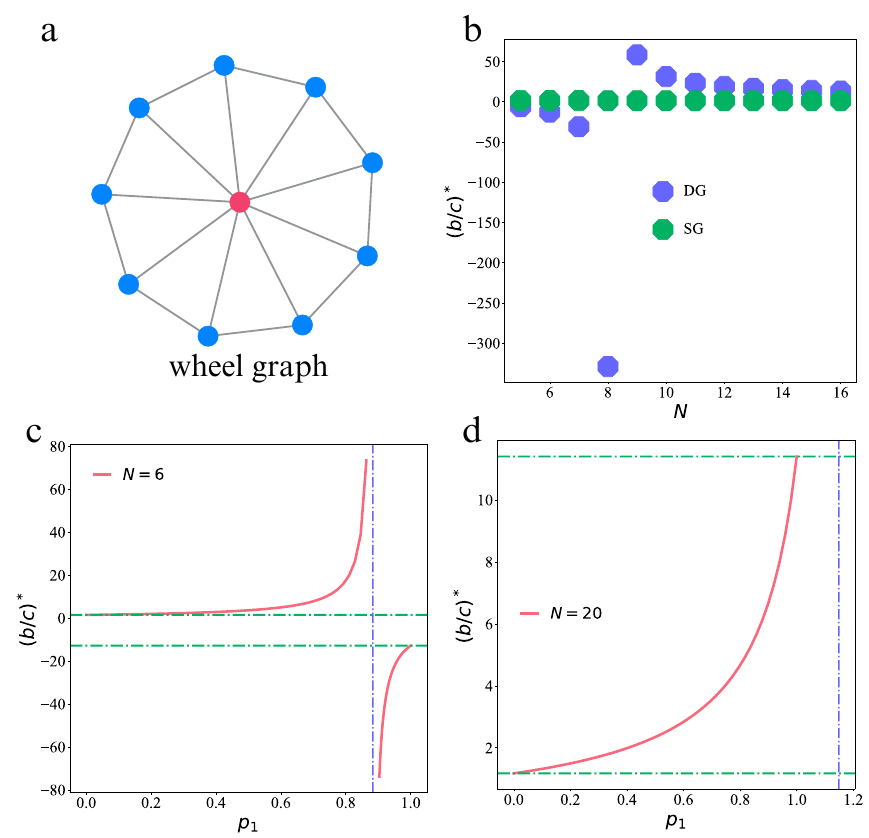}
\caption{\textbf{Effects of exogenous game transitions with heterogeneity on wheel graphs.} We consider wheel graphs for further analysis, as shown in subplot~(a). Subplot~(b) shows how $(b/c)^*$ varies with graph size $N$ under DG and SG. When $(b/c)^*_\text{DG} < 0$, we can observe an analogous result in complete networks (subplot~(c)). When $(b/c)^*_\text{DG} > 0$, $(b/c)^*_\text{SG}$ presents continuous changes (subplot~(d)). It is similar to the outcomes in star graphs, except it has an upper bound.}
\end{figure}

\clearpage

\begin{figure}
\centering
\includegraphics[width=\textwidth]{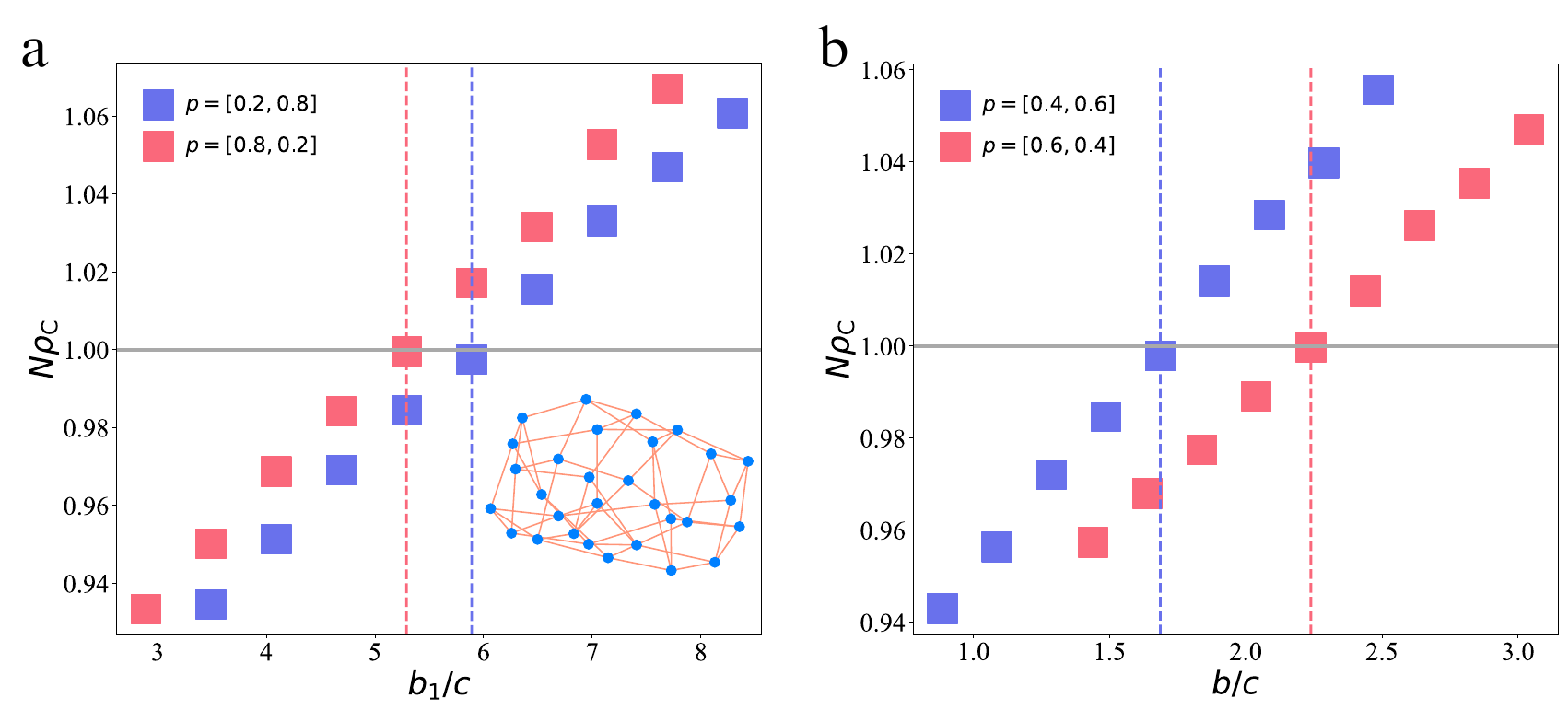}
\caption{\textbf{Evolutionary dynamics with exogenous game transitions under weak selection.} Fixation probabilities are averaged over $10^7$ independent runs with selection intensity $\delta=0.01$ and the cost of cooperative behavior $c=1$. The vertical lines in each subplot show the theoretical values for the critical benefit-to-cost ratio. The population structure is a random regular graph with size $N=30$ and degree $k=4$. Subplot~(a) (resp. (b)) exhibits the exogenous game transitions with homogeneity (resp. heterogeneity). In subplot~(b), the appearance of SG effectively reduces the threshold for the dominance of cooperation. When heterogeneity is taken into account (subplot~(b)), all game states are initialized as snowdrift games, whereas in the absence of heterogeneity (subplot~(a)), they are less valuable donation games. Simulations agree well with the theoretical predictions. } 
\end{figure}

\clearpage

\begin{figure}
    \centering
    \includegraphics[width=\textwidth]{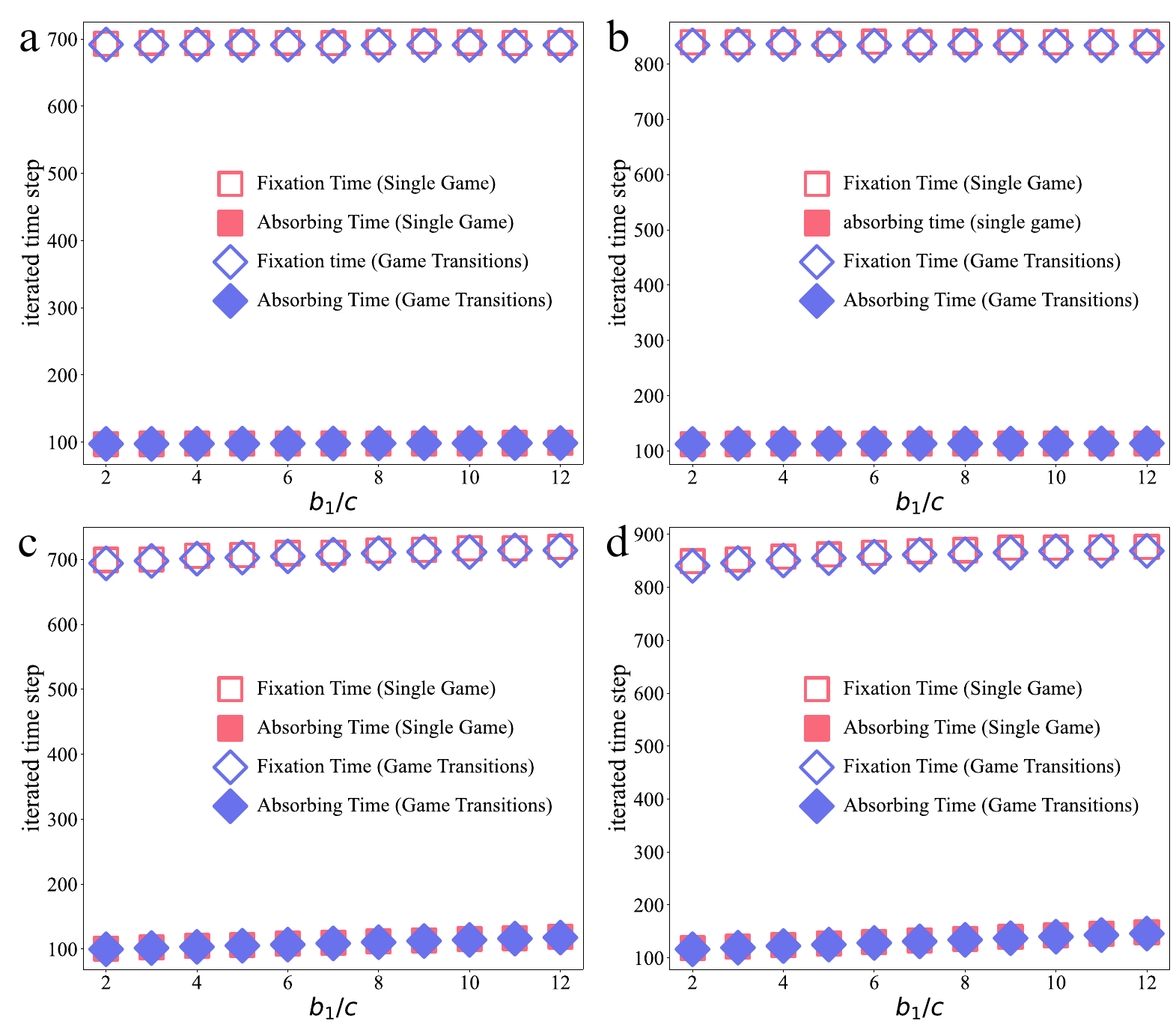}
    \caption{\textbf{Deterministic game transition has no effects on fixation time and absorbing time.} The absorbing time means the average time to reach one of the absorbing states (all defectors or all cooperators), and the fixation time describes the average time to reach the state full of cooperators. We present the iterated time as a function of the benefit-to-cost ratio on the Barab\'{a}si-Albert graph (subplots (a) and (c)) and Erd\H{o}s-R\'{e}nyi graph (subplots (b) and (d)) and in the donation game (subplots (a) and (b)) and snowdrift game (subplots (c) and (d)).}
\end{figure}

\clearpage

\begin{figure}
    \centering
    \includegraphics[width=\textwidth]{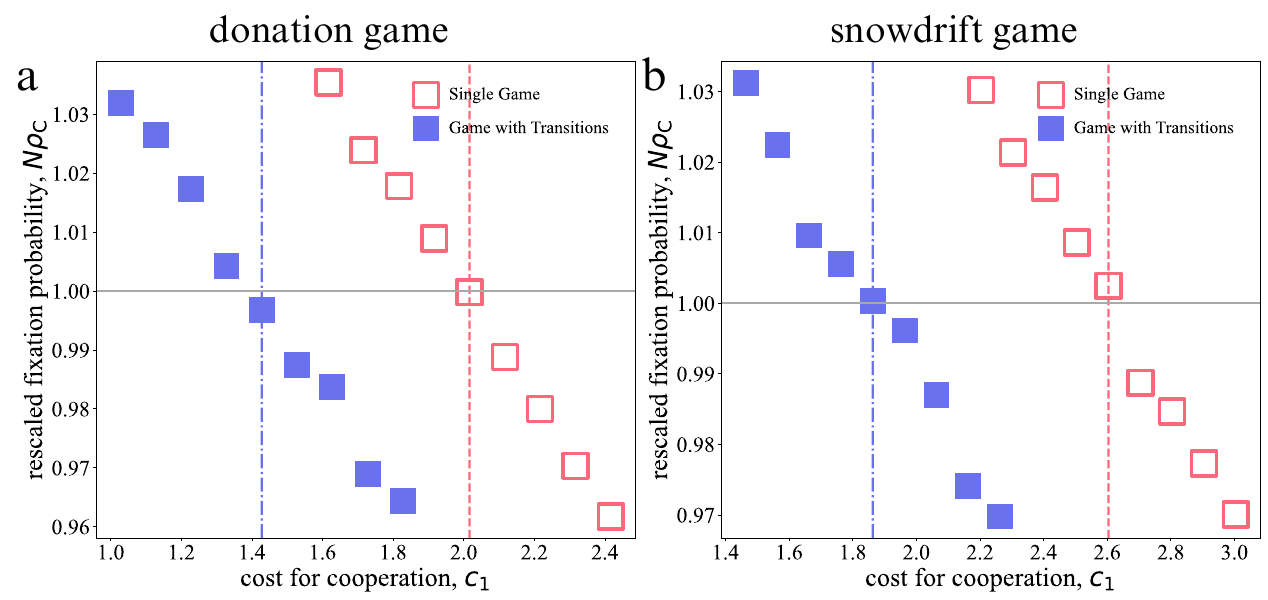}
    \caption{\textbf{Effects on dynamically adjusting costs on evolutionary outcomes under death-birth updating.} We present the rescaled fixation probability of cooperation, $N\rho_{\textrm{C}}$, as a function of the cost $c_1$ on Barab\'{a}si-Albert (BA) graphs of size $N$ in the donation game (left panel) and the snowdrift game (right panel). Cooperation is favored if $N\rho_\textrm{C}$ exceeds $1$ (horizontal lines), i.e., $\rho_{\textrm{C}}>1/N$. Vertical lines represent corresponding analytical predictions of the critical benefit-to-cost ratios $(c_1)^*$, below which the evolution of cooperation is facilitated. The cost difference $\Delta c = c_1-c_2 = -1$ is the same in the two games. The benefits of the donation game and snowdrift game are $b=20$ and $b=3$, respectively. Other parameters are the same in Fig.~3.}
\end{figure}

\clearpage

\begin{figure}
\centering
\includegraphics[width=\textwidth]{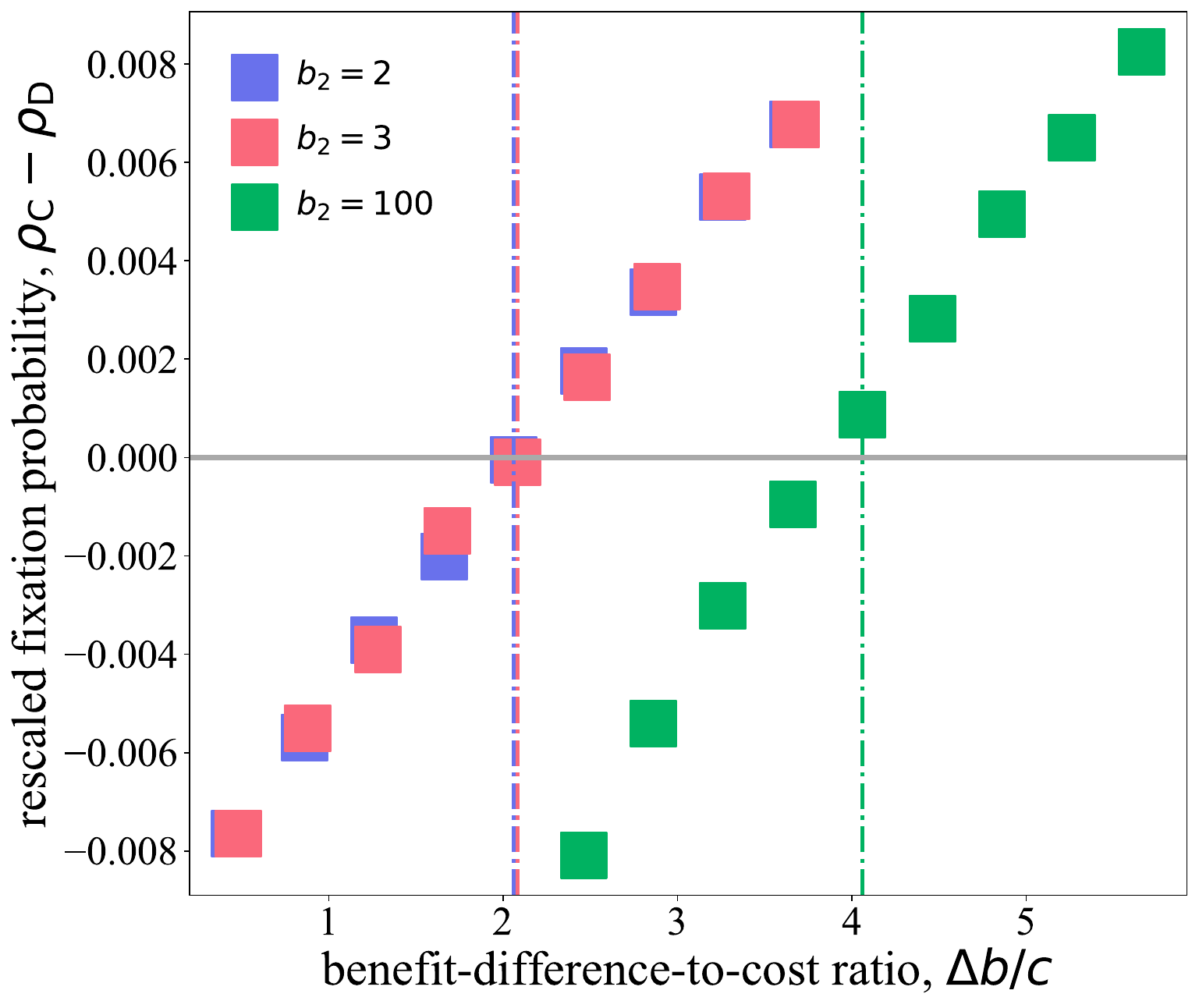}
\caption{\textbf{Effects of game transitions on evolutionary dynamics under pairwise-comparison updating.} We present the fixation probability of cooperation versus defection, $\rho_{\textrm{C}} - \rho_{\textrm{D}}$, as a function of the ratio $\Delta b/c$ in the donation game under different $b_2$ on the random regular graph with size $N=100$ and degree $k=4$. Cooperation is favored if $\rho_\textrm{C} - \rho_\textrm{D}$ is greater than $0$ (horizontal lines), i.e., $\rho_{\textrm{C}}>\rho_{\textrm{D}}$. Vertical dash-dotted lines correspond to required critical ratios $(\Delta b/c)^*$ for the dominance of cooperation. For small $b_2$ compared to size $N$, $(\Delta b/c)^* \approx 2$, and while that is not the case for larger $b_2$ (e.g., $b_2 = 100$ in the green curve).}
\end{figure}

\clearpage

\begin{figure}
    \centering
    \includegraphics[width=\textwidth]{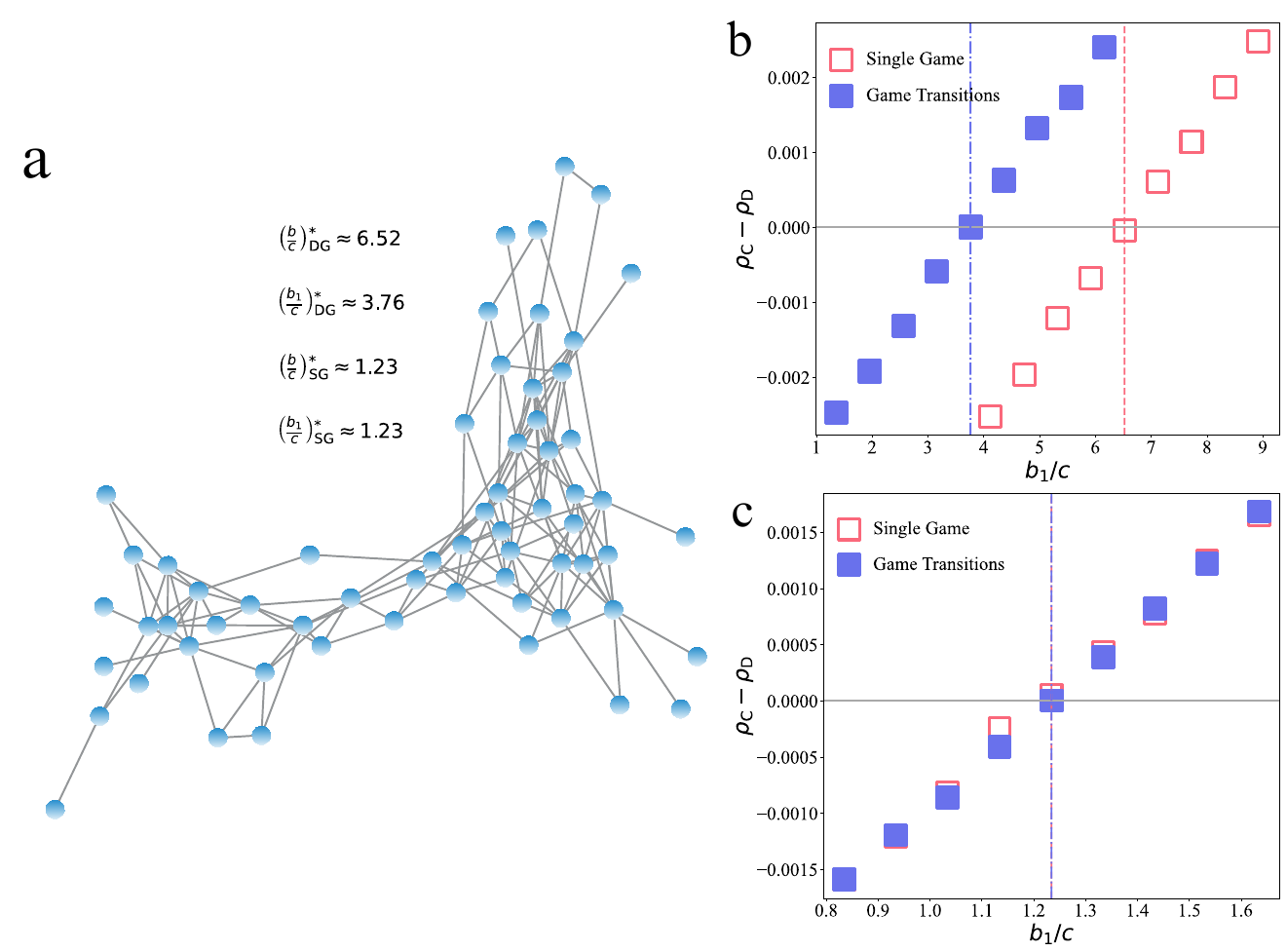}
    \caption{\textbf{Simulations on the social network with game transitions.} We present the fixation probability of cooperation against defection, $\rho_{\textrm{C}} - \rho_{\textrm{D}}$, as a function of the benefit-to-cost ratio $(b_1/c)$ in the donation game (subplot (b)) and snowdrift game (subplot (c)) on the empirical network from dolphins' associations, of which the structure is shown in subplot (a). The network statistics of the population structure are provided in Tab.~1 in the main text. Cooperation is facilitated if $\rho_\textrm{C}-\rho_\textrm{D}$ exceeds $0$ (horizontal lines), i.e., $\rho_{\textrm{C}}>\rho_\textrm{D}$. The approximate value of each critical benefit-to-cost ratio is given in subplot (a). Other settings are the same in Fig.~3.}
\end{figure}

\clearpage

\begin{figure}
\centering
\includegraphics[width=\textwidth]{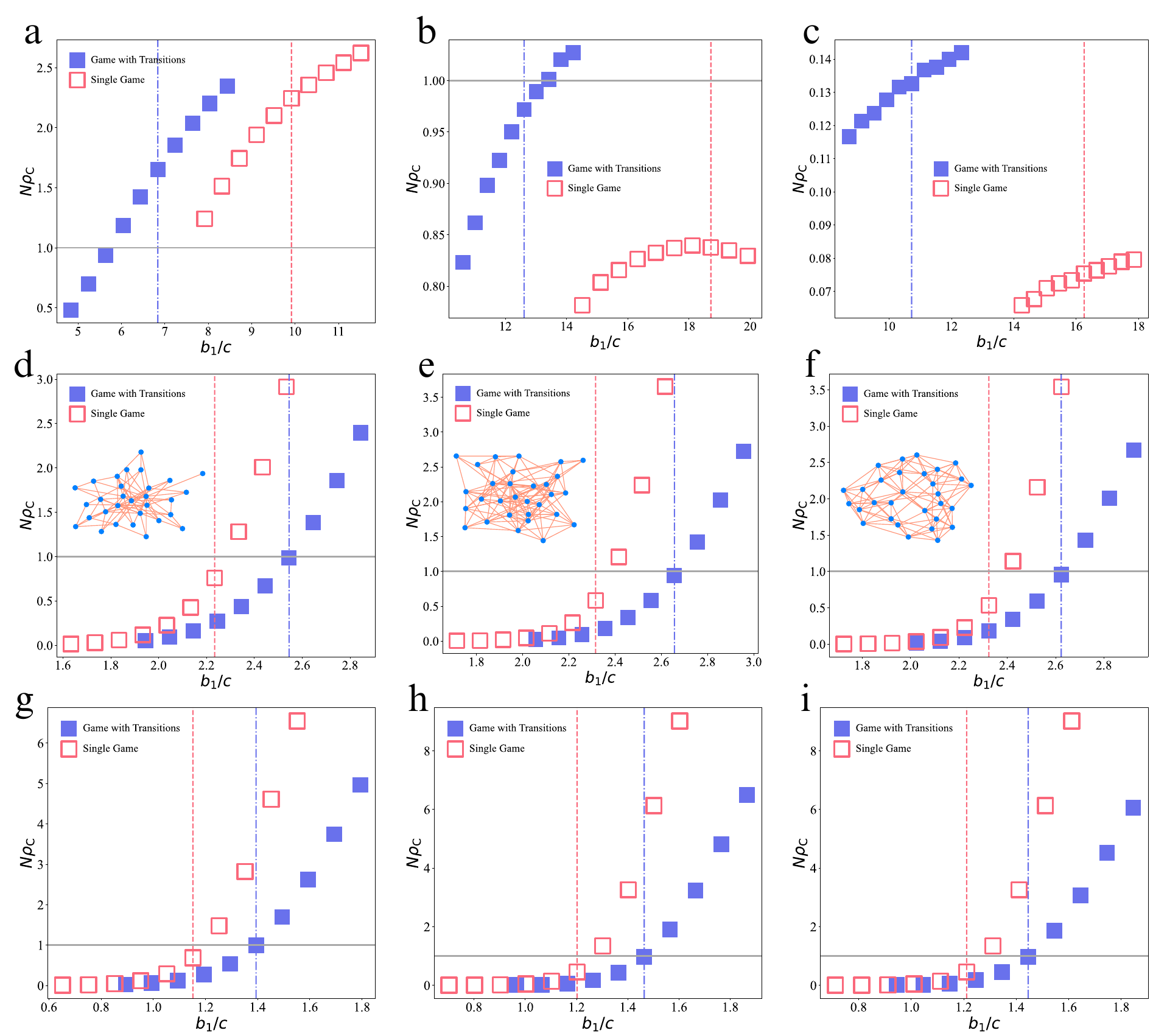}
\caption{\textbf{Effects of game transitions on evolutionary dynamics under strong selection and death-birth updating.} We set $\delta = 1$ and $F_i(\mathbf{s},\mathbf{g}) = \exp(\delta f_i(\mathbf{s},\mathbf{g}))$, which is equivalent to $F_i(\mathbf{s},\mathbf{g}) = 1 + \delta f_i(\mathbf{s},\mathbf{g})$ used in the main text under weak selection. Evolutionary outcomes are increasingly complicated under strong selection. Game transitions cannot allow the dominance of cooperation in all topologies (subplot~(c)). Fixation probabilities are obtained by the same Monte Carlo method applied in Fig.~2 in the main text.}
\end{figure}

\clearpage

\begin{figure}
\centering
\includegraphics[width=\textwidth]{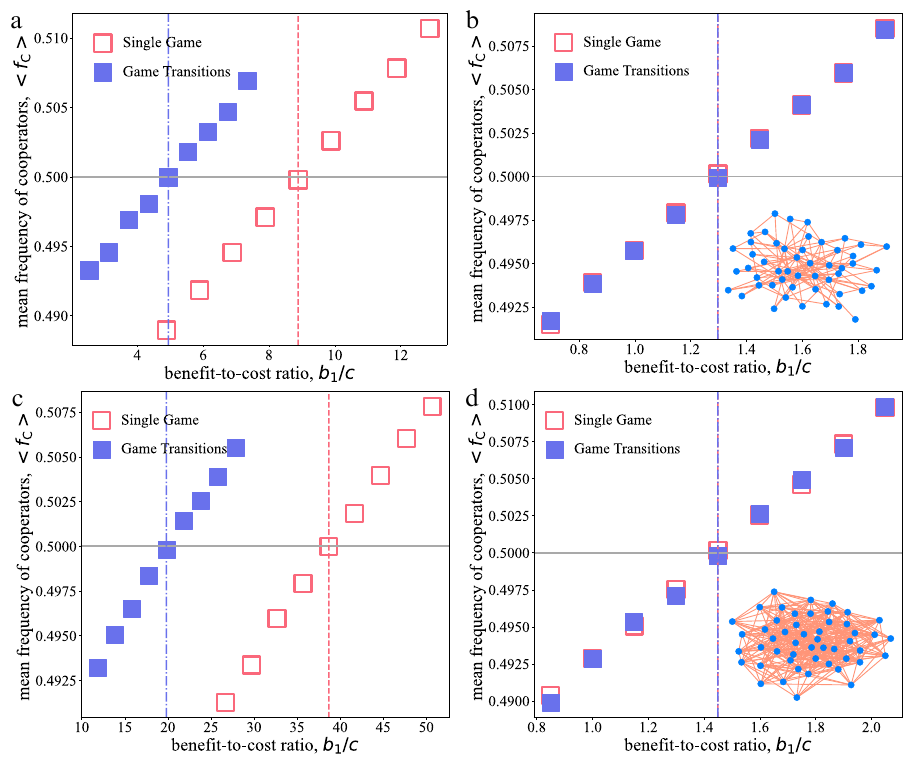}
\caption{\textbf{Effects of game transitions on cooperative behaviors in the presence of mutation under death-birth updating.} Two social dilemmas: donation games (subplots~(a) and~(c)) and snowdrift games (subplots~(b) and~(d)), and two social structures with size $N=50$: Barab\'{a}si-Albert (subplots~(a) and~(b)) and Erd\H{o}s-R\'{e}nyi (subplots~(c) and~(d)) graphs, are considered. The mutation rate $u$ ($u = 0.05$ in simulations) refers to with probability $1-u$, the dead node is replaced by its neighbor's offspring, and with probability $u$, the mutation transfers it to a cooperator or a defector with uniform probability. Mean frequency of cooperators $\left<f_{\textrm{C}}\right>$ is simulated by averaging over $10^8$ generations in $50$ independent runs. Each node has an equal probability of becoming a cooperator or defector in the initial stage. Other parameters are the same in Fig.~2.}
\end{figure}

\clearpage

\bibliography{sn-bibliography}

\end{document}